\documentclass[
  aps,
  prb,
  showpacs,
  superscriptaddress,
  floatfix,
  twocolumn,
  secnumarabic,
  amssymb,
  amsmath]{revtex4-1}

\usepackage{graphicx}
\usepackage{bm}
\usepackage{color}
\usepackage[colorlinks=true,citecolor=blue,linkcolor=blue,pdfstartview=FitH,pdfauthor=Gepraegs]{hyperref}

\begin{document}

\preprint{Gepraegs \textit{et al.}, version: 2012-08-28}

\title{Giant Magnetoelastic Effects in BaTiO$_{3}$-based Extrinsic Multiferroic Hybrids}

\author{Stephan Gepr\"{a}gs}
 \email{Stephan.Gepraegs@wmi.badw.de}
 \affiliation{Walther-Mei{\ss}ner-Institut, Bayerische Akademie der
              Wissenschaften, 85748 Garching, Germany}

\author{Matthias Opel}
 \affiliation{Walther-Mei{\ss}ner-Institut, Bayerische Akademie der
              Wissenschaften, 85748 Garching, Germany}

\author{Sebastian T. B.~Goennenwein}
 \affiliation{Walther-Mei{\ss}ner-Institut, Bayerische Akademie der
              Wissenschaften, 85748 Garching, Germany}

\author{Rudolf Gross}
 \email{Rudolf.Gross@wmi.badw.de}
 \affiliation{Walther-Mei{\ss}ner-Institut, Bayerische Akademie der
              Wissenschaften, 85748 Garching, Germany}
 \affiliation{Physik-Department, Technische Universit\"{a}t M\"{u}nchen, 85748 Garching, Germany}

\date{\today}%

\begin{abstract}
Extrinsic multiferroic hybrid structures consisting of ferromagnetic and ferroelectric layers elastically coupled to each other are promising due to their robust magnetoelectric effects even at room temperature. For a quantitative analysis of these magnetoelectric effects, a detailed knowledge of the piezoelectric and magnetoelastic behavior of both constituents as well as their mutual elastic coupling is mandatory. We here report on a theoretical and experimental study of the magnetic behavior of BaTiO$_{3}$-based extrinsic multiferroic structures. An excellent agreement between molecular dynamics simulations and the experiments was found for Fe$_{50}$Co$_{50}$/BaTiO$_{3}$ and Ni/BaTiO$_{3}$ hybrid structures. This demonstrates that the magnetic behavior of extrinsic multiferroic hybrid structures can be determined by means of ab-initio calculations, allowing for the design of novel multiferroic hybrids.
\end{abstract}


\maketitle

\section{Introduction}
\label{sec:intro}

Multiferroic materials,\cite{Schmid:162:1994} which simultaneously possess at least two long-range ordering phenomena in the same phase, have attracted a lot of attention in recent years due to their rich physics and large variety of potential applications.\cite{Eerenstein:442:2006,Ramesh:6:2007} Moreover, the mutual coupling between the ferroic order parameters lays the foundations for a variety of novel phenomena as well as enhanced functionalities and improved properties in future engineered material systems.\cite{Spaldin:309:2005,Gepraegs:87:2007} In this field, so-called extrinsic multiferroic hybrid structures, in which ferromagnetic and ferroelectric compounds are artificially assembled, are promising candidates. They enable large and robust cross-coupling effects at room temperature by exploiting the elastic coupling between the two constituents, leading to extrinsic magnetoelectric effects.\cite{Zheng:303:2004,Eerenstein:6:2007,Nan:103:2008,Srinivasan:40:2010} Such structures can be realized in the form of particular composites,\cite{Run:9:1974} laminate composites,\cite{Ryu:40:2001} vertical nanostructures,\cite{Zheng:303:2004} and horizontal hybrid structures.\cite{Murugavel:85:2004} Concentrating on horizontal hybrid structures consisting of ferromagnetic and ferroelectric layers, various materials were used to realize strain-mediated multiferroic horizontal hybrids in the past years.\cite{Vaz:22:2010} Even industrially produced piezoelectric actuators \cite{Brandlmaier:77:2008,Bihler:78:2008,Weiler:11:2009,Brandlmaier:110:2011} or multilayer capacitors \cite{Israel:7:2008} were exploited to achieve extrinsic magnetoelectric effects.

In the context of magnetoelectric effects, the electric control of magnetism is of particular interest. This converse magnetoelectric coupling relies on mechanical deformations of the ferroelectric layer caused by the converse piezoelectric effect, or by ferroelectric/ferroelastic domain reconfigurations. These elastic strain changes are transferred into the ferromagnetic thin film clamped onto the ferroelectric layer, modifying its magnetic properties due to magnetoelastic effects.\cite{Ma:23:2011} In total, converse magnetoelectric effects in multiferroic hybrids can be described phenomenologically by a product tensor property including piezoelectric and mangetoelastic effects.\cite{Suchtelen:27:1972,Nan:50:1994} Thus, a detailed understanding of these effects and the elastic coupling across the interface is mandatory to predict magnetoelectric effects in novel multiferroic hybrid structures.

As a first step, we here investigate magnetization changes caused by magnetoelastic effects in BaTiO$_{3}$ (BTO) based extrinsic multiferroic hybrid structures using both Fe$_{50}$Co$_{50}$ (FeCo) and Ni as ferromagnetic materials. The experimental approach is similar to Lee \textit{et al.}.\cite{Lee:77:2000} They reported large changes of the magnetization and the electric resistivity in La$_{0.7}$Sr$_{0.3}$MnO$_{3}$/BaTiO$_{3}$ (LSMO/BTO) extrinsic multiferroic hybrids, whenever BTO undergoes a structural phase transition. In recent years, further experiments using LSMO/BTO,\cite{Dale:82:2003,Eerenstein:6:2007} Fe/BTO,\cite{Sahoo:76:2007,Taniyama:105:2009,Brivio:98:2011} Fe$_{3}$O$_{4}$/BTO,\cite{Tian:92:2008,Vaz:94:2009,Sterbinsky:96:2010} CoFe$_{2}$O$_{4}$/BTO,\cite{Chopdekar:89:2006} and Sr$_{2}$CrReO$_{6}$/BTO\cite{Czeschka:95:2009,Gepraegs:321:2009,Opel:208:2011,Opel:45:2012} hybrid structures were performed. In these experiments, magnetization jumps were observed caused by strain changes of the respective ferromagnetic thin film, which are induced by the structural phase transitions of BTO. However, the experimental results could hardly be explained in a quantitative manner, since ferroelectric BTO forms a multi-domain state upon crossing the phase transitions without any external field applied to the crystal. This nonuniform strain state is induced into the overlying ferromagnetic thin film and causes changes of its magnetic properties. In particular, since the volume fraction of the domains in each ferroelastic phase of BTO may not exhibit the same value upon cooling down or warming up the hybrid structure, a different behavior of the overlying ferromagnetic thin film was observed during decreasing and increasing temperature.\cite{Chopdekar:89:2006,Sahoo:76:2007,Czeschka:95:2009} This demonstrates, it is essential to control both the ferromagnetic and ferroelastic domain configuration in order to correctly describe the manipulation of the magnetization in BTO-based multiferroic hybrid structures.

By using miscut BTO substrates, we show that the magnetization changes of multiferroic hybrid structures can indeed be predicted on the basis of first-principles effective Hamiltonian simulations. To calculate these changes, we first determine the strain state of miscut BTO crystals as a function of temperature by means of molecular dynamics (MD) simulations in Section~\ref{sec:BTO}. Since the overlying ferromagnetic thin film is clamped onto the BTO substrate, each change of the in-plane strain state of the BTO crystal modifies the strain state of the ferromagnetic thin film. Second, the strain state of the ferromagnetic thin film is determined under the assumption of an ideal strain transfer between the BTO crystal and the ferromagnetic thin film in Section~\ref{sec:FMstrain}. Knowing the strain state of the ferromagnetic thin film, we then show in Section~\ref{sec:MTtheory} that the magnetization can be calculated using a phenomenological thermodynamic model. To compare these calculations to experimental results, we fabricated FeCo/BTO and Ni/BTO hybrid structures and investigated their magnetic properties by SQUID magnetometry. In Section~\ref{sec:experiment} we show that there is excellent agreement between the experimental results and the theoretical simulations. This demonstrates that our approach allows us to predict the magnetic behavior of existing and novel ferromagnetic/ferroelectric hybrid structures based on first-principles calculations.

\section{BaTiO$_{3}$ as the ferroelectric material}
\label{sec:BTO}

The use of BTO as the ferroelectric constituent allows us to modify the magnetic properties of the overlying ferromagnetic thin film not only by exploiting piezoelectric effects at constant temperature but also by using strain effects due to changes of the in-plane lattice constant, which occur as a function of temperature at the natural phase transitions of BTO. Starting in the cubic paraelectric state ($T>T_{\mathrm{c}}\simeq 393$\,K),\cite{Sakayori:JJAP:34:1995} BTO undergoes successive structural phase transitions to three different ferroelectric phases as the temperature is lowered.\cite{Merz:76:1949} Each transition is accompanied by a change of the crystallographic symmetry, which causes large changes of the strain state of the ferromagnetic film clamped onto the BTO substrate.\cite{Shebanov:65:1981} The first transition occurs at the ferroelectric Curie temperature $T_{\mathrm C}$, at which the lattice symmetry of BTO changes from cubic to tetragonal. In the tetragonal phase, the polar axis is aligned along one of the pseudo-cubic $\langle 100 \rangle_{\mathrm{pc}}$ directions. Since there are six equivalent $\langle 100 \rangle_{\mathrm{pc}}$ axes, six ferroelectric and three ferroelastic domains are possible. These domains are denoted as $c$- or $a$-domains depending on weather the polarization is pointing out-of-plane along the $[001]_{\mathrm{pc}}$ direction or along the in-plane directions $[100]_{\mathrm{pc}}$ and $[010]_{\mathrm{pc}}$, respectively. Within the ferroelectric state, two more phase transitions occur. At $T \simeq 278$\,K, the lattice symmetry changes into an orthorhombic structure and at $T \simeq 183$\,K, it is further reduced to rhombohedral. At both phase transitions, a reorientation of the polar axis takes place. In the orthorhombic (rhombohedral) state, the polar axis is aligned along one of the pseudo-cubic $\langle 110 \rangle_{\mathrm{pc}}$ ($\langle 111 \rangle_{\mathrm{pc}}$) directions. Since the parent, paraelectric structure of BTO is cubic, it is convenient to describe the lattice structure of BTO as a distorted cubic structure in the whole temperature range.

\begin{figure}[tb]
  \includegraphics[width=\columnwidth]{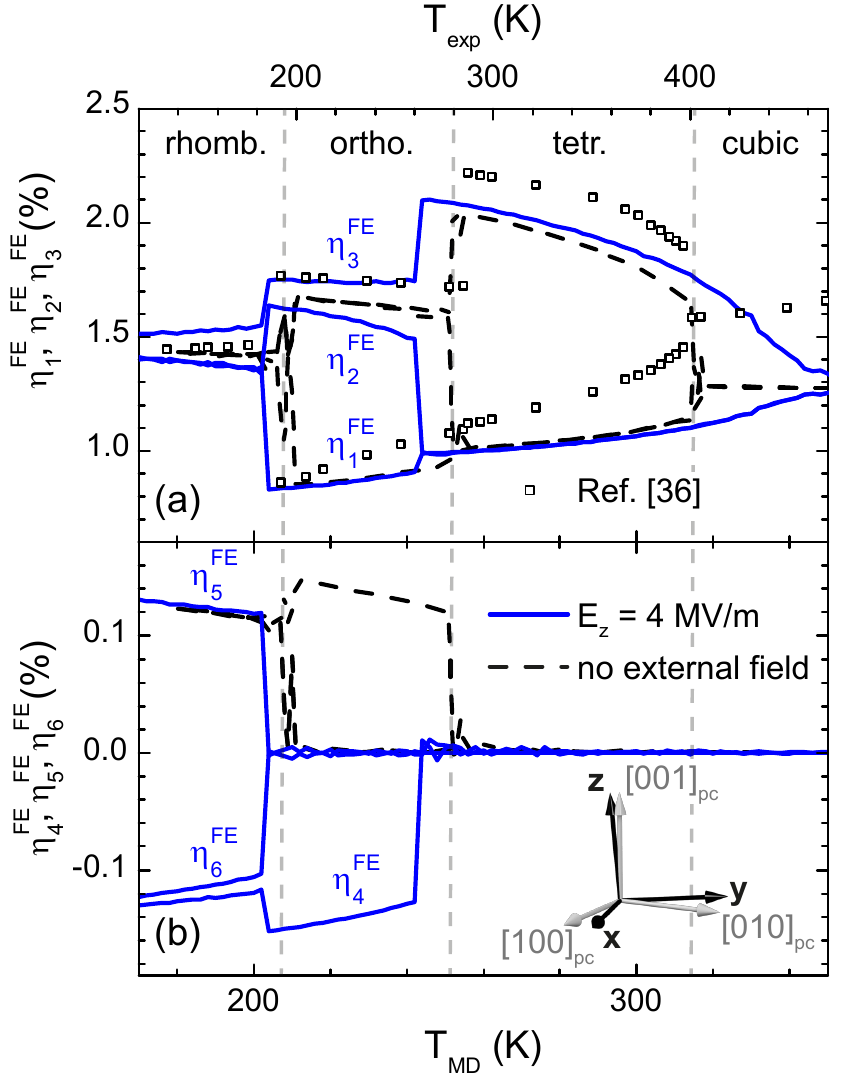}
    \caption{(color online)
Average homogeneous strain components $\eta^{\mathrm{FE}}_{k}$ with $k=1,\ldots,6$ as a function of temperature $T_{\mathrm{MD}}$ calculated by performing MD simulations [cf.~Ref.~\onlinecite{Nishimatsu:78:2008}]. The evolution of the absolute value of $\eta^{\mathrm{FE}}_{k}$ simulated during cooling from 420\,K without any external field applied is marked by dashed lines. For comparison, open symbols represent experimental data as a function of $T_{\mathrm{exp}}$ [adapted from Ref.~\onlinecite{Shebanov:65:1981}]. The strains are calculated relative to the lattice constant of $a=0.3948$\,nm derived by LDA calculations. The temperature of the phase transitions are indicated by vertical dashed lines. To reduce the number of possible ferroelastic domains in each phase, MD simulations are performed with an electric field of 4\,MV/m applied along the $\mathbf{z}$-direction and assuming a miscut BTO crystal (solid blue lines). The difference between the pseudo-cubic $[001]_{\mathrm{pc}}$-direction ($[100]_{\mathrm{pc}}$-direction) and the $\mathbf{z}$-direction ($\mathbf{x}$-direction) was assumed to be $1^\circ$ ($7^\circ$).}
    \label{fig:Fig1}
\end{figure}

By employing the strain changes at the natural phase transitions of BTO, large modifications of the magnetization are expected in multiferroic ferromagnet/BTO hybrid structures.\cite{Lee:77:2000} To quantitatively calculate these changes, we first determine the strain state of BTO single crystals by performing MD simulations using the \textit{FERAM} code developed by Takeshi Nishimatsu.\cite{FERAMcode} This simulation is based on a first-principles effective Hamiltonian constructed from local-density approximation (LDA) calculations and described in detail in Refs.~\onlinecite{Nishimatsu:78:2008,Nishimatsu:82:2010}. For the calculations, we used a $16 \times 16 \times 16$ supercell with periodic boundary conditions.\cite{Paul:99:2007} Furthermore, the system was thermalized within 60~000 time steps, and the properties were averaged over 60~000 steps. Since first-principles density-functional-theory (DFT) calculations normally underestimate lattice constants, a negative pressure of $p=-5.0$\,GPa was used in the MD simulations.\cite{Nishimatsu:78:2008} Since the ferromagnetic film in our experiments is more than 1000 times thinner than the ferroelectric BTO substrate, we can safely assume that the strain state of the ferromagnet/BTO hybrid structure is only determined by the BTO component. 

In Fig.~\ref{fig:Fig1}, the absolute values of the average homogeneous strain components $\eta^{\mathrm{FE}}_{k}$ with $k=1,...,6$ (in matrix notation: $\eta_{1}=\epsilon_{11}$, $\eta_{2}=\epsilon_{22}$, $\eta_{3}=\epsilon_{33}$, $\eta_{4}=2\epsilon_{23}$, $\eta_{5}=2\epsilon_{31}$, $\eta_{6}=2\epsilon_{12}$) of BTO are simulated as a function of temperature $T_{\mathrm{MD}}$ during cooling from 420\,K without any external fields. The strains are calculated relative to the equilibrium cubic lattice structure with a lattice constant of $a=0.3948$\,nm derived by LDA calculations.\cite{Nishimatsu:78:2008} To compare the simulation with experimental data, literature values published by Shebanov \textit{et al.}\cite{Shebanov:65:1981} are included [cf.~open symbols in Fig.~\ref{fig:Fig1}(a)]. As obvious from Fig.~\ref{fig:Fig1}, the MD calculations reveal the correct sequence of phase transitions of BTO. However, in spite of the negative pressure applied, the transition temperatures are still underestimated and the temperature scales of the MD calculations $T_{\mathrm{MD}}$ and the experiment $T_{\mathrm{exp}}$ do not coincide. This is usually corrected by linearly rescaling the temperature axis in order to adjust the theoretical and experimental phase transition temperatures.\cite{Garcia:72:2981} Moreover, the MD calculations reproduce the experimental behavior only for temperatures $T_{\mathrm{exp}} \leq 300$\,K. The discrepancy at $T_{\mathrm{exp}} > 300$\,K can mainly be attributed to the poor description of thermal expansion effects in the effective Hamiltonian as well as to the underestimation of the lattice constants in DFT calculations.\cite{Nishimatsu:82:2010} Nevertheless, upon rescaling $T_{\mathrm{MD}}$, the elastic behavior of BTO can be well described by MD calculations for $T_{\mathrm{exp}} \leq 300$\,K.

\begin{figure}[tb]
  \includegraphics[width=\columnwidth]{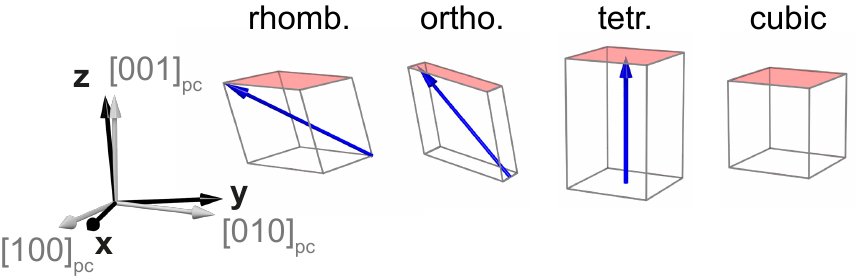}
    \caption{(color online)
Schematic representation of the energetically most favorable ferroelastic domains in the rhombohedral, orthorhombic, tetragonal and cubic phase under an electric field applied along the $\mathbf{z}$-direction using miscut BTO crystals. The blue arrows indicate the orientation of the ferroelectric polarization. The polarization rotation induced by the electric field is neglected for simplicity.}
    \label{fig:Fig2}
\end{figure}

It is important to note that, without applying external fields, a multi-domain state is expected in BTO in all three ferroelectric/ferroelastic phases upon cooling, since the different ferroelectric polarization directions in each phase are energetically degenerated and have the same probability to appear.\cite{Catalan:84:2012} As the formation of ferroelastic domains depends strongly on extrinsic effects, such as sample shape, structural defects and imperfections,\cite{Damjanovic:61:1998} the simulation of the detailed strain state in each phase is hardly possible. To resolve this issue and to allow an unambiguous comparison of experiment and simulation results, a full control of the ferroelastic domain configuration, which leads to a well defined strain state in the BTO crystal, is mandatory.\cite{Park:86:1999} This can be achieved by using miscut BTO crystals. In this case, the coordinate system describing the ferroelastic domains in a cubic reference system by $[100]_{\mathrm{pc}}$, $[010]_{\mathrm{pc}}$, and $[001]_{\mathrm{pc}}$ differs from the directions $\mathbf{x}$, $\mathbf{y}$, and $\mathbf{z}$ describing the surface of the BTO crystal [cf.~inset of Fig.~\ref{fig:Fig1}(b)]. By additionally applying an external electric field along the $\mathbf{z}$-direction, one ferroelastic domain becomes energetically favorable in each ferroelastic phase. The resulting ferroelastic domain evolution is schematically shown in Fig.~\ref{fig:Fig2}, where any polarization rotations induced by the applied electric field\cite{Fu:403:2000} are neglected for simplicity.
%
%
This situation can be simulated by performing MD calculations with an electric field of $E_{\mathrm{z}}=4$\,MV/m, which is theoretically needed to ensure a single domain state in the whole temperature range. The difference between the pseudo-cubic $[001]_{\mathrm{pc}}$ ($[100]_{\mathrm{pc}}$) axis and the $\mathbf{z}$-direction ($\mathbf{x}$-direction) was assumed to be $1^\circ$ ($7^\circ$). Figure~\ref{fig:Fig1} reveals that the cubic to tetragonal phase transition becomes diffuse and the transition temperature increases by about 40\,K (cf.~solid line in Fig.~\ref{fig:Fig1}). A slight decrease of the transition temperatures between the tetragonal and orthorhombic phase as well as between the orthorhombic and rhombohedral phase can also be observed. Furthermore, the symmetry of the orthorhombic and rhombohedral phases are reduced.\cite{Vanderbilt:63:2001,Paul:80:2009} In spite of these symmetry changes, both phases are consistently labeled by the parent orthorhombic and rhombohedral phases throughout this paper. As shown in Fig.~\ref{fig:Fig1}, every strain component $\eta^{\mathrm{FE}}_{k}$ is uniquely defined in the whole temperature range. Thus, with the knowledge of the deformation of miscut BTO crystal, the magnetization changes of a ferromagnetic thin film deposited on top can now be calculated employing magnetoelastic theory. To this end, the elastic behavior of the ferromagnetic thin film is determined first.

\section{Strain state of the ferromagnetic thin film}
\label{sec:FMstrain}

In so-called horizontal multiferroic hybrid structures, a ferromagnetic thin film is elastically coupled to a usually much thicker ferroelectric substrate. Neglecting surface effects and assuming a perfect elastic coupling between both constituents, the strain state of the ferromagnetic thin film can be considered as well-defined and homogeneous throughout the entire volume. Since the ferromagnetic thin film is clamped to the thick ferroelectric substrate, its in-plane strain components $\eta^{\mathrm{FM}}_{1}$, $\eta^{\mathrm{FM}}_{2}$, and $\eta^{\mathrm{FM}}_{6}$ are totally controlled by the ferroelectric substrate: $\eta^{\mathrm{FM}}_{1}=\eta^{\mathrm{FE}}_{1}$, $\eta^{\mathrm{FM}}_{2}=\eta^{\mathrm{FE}}_{2}$, $\eta^{\mathrm{FM}}_{6}=\eta^{\mathrm{FE}}_{6}$. Thus, only the remaining components $\eta^{\mathrm{FM}}_{3}$, $\eta^{\mathrm{FM}}_{4}$, and $\eta^{\mathrm{FM}}_{5}$ have to be calculated to determine the total strain state of the ferromagnetic thin film. This situation is equal to the general case of pseudomorphic or coherent growth of epitaxial thin films on crystalline substrates.\cite{Marcus:55:1994} By using a phenomenological thermodynamic model the strain components can be calculated employing the fact that the out-of-plane stress components $\sigma^{\mathrm{FM}}_{3}$, $\sigma^{\mathrm{FM}}_{4}$, and $\sigma^{\mathrm{FM}}_{5}$  are zero, since no forces are acting on the free surface of the ferromagnetic thin film.\cite{Brandlmaier:77:2008} In the early work of Pertsev \textit{et al.},\cite{Pertsev:80:1998} it was shown that due to mechanical boundary conditions the equilibrium thermodynamic state of a thin film clamped to a rigid substrate is described by the thermodynamic potential $\tilde{g}^{\mathrm{FM}}$, which is given by the Legendre transformation of the well known Gibbs free energy density $g^{\mathrm{FM}}$ by\cite{Pertsev:80:1998} 

\begin{equation}
   \tilde{g}^{\mathrm{FM}} = g^{\mathrm{FM}}+\sigma^{\mathrm{FM}}_{1}\eta^{\mathrm{FM}}_{1}+\sigma^{\mathrm{FM}}_{2}\eta^{\mathrm{FM}}_{2}+\sigma^{\mathrm{FM}}_{6}\eta^{\mathrm{FM}}_{6} \; .
\label{eq:free_enthalpy}
\end{equation}

Thus, the strain components $\eta^{\mathrm{FM}}_{k}$ with $k=3,\,4,\,5$ are derived from $\sigma^{\mathrm{FM}}_{k}=\partial \tilde{g}^{\mathrm{FM}}/\eta^{\mathrm{FM}}_{k}=0$. To determine $\tilde{g}^{\mathrm{FM}}$, the magnetic energy density $u^{\mathrm{FM}}$ of the ferromagnetic thin film, which depends on the magnetization $M_{i}$ and the strain state $\eta^{\mathrm{FM}}_{k}$, is calculated. In the following we restrict our discussion to situations where the magnetization $\mathbf{M}=M_{\mathrm{s}}\mathbf{m}$ is well-defined and the unit vector $\mathbf{m}$ as well as the saturation magnetization $M_{\mathrm{s}}$ is homogeneous throughout the thin film. This is fulfilled for temperatures well below the magnetic Curie temperature in the absence of magnetic domains. In this Stoner-Wohlfarth type of approach,\cite{Stoner:240:1948} the direction of the magnetization can be expressed by the components of $\mathbf{m}$,~i.e., by the directional cosines $m_{i}$. To take into account magnetoelastic effects, $u^{\mathrm{FM}}(m_{i},\eta^{\mathrm{FM}}_{k})$ can be expanded in powers of $m_{i}$ and $\eta^{\mathrm{FM}}_{k}$ for small and homogeneous deformations.\cite{OHandley:book} Thus, $u^{\mathrm{FM}}=u^{\mathrm{FM}}(m_{i},\eta^{\mathrm{FM}}_{k})-u_{0}$ contains the lowest-order terms of three contributions:

\begin{equation}
   u^{\mathrm{FM}}=u^{\mathrm{FM}}_{\mathrm{ani}}(m_{i})+u^{\mathrm{FM}}_{\mathrm{el}}(\eta^{\mathrm{FM}}_{k})+u^{\mathrm{FM}}_{\mathrm{magel}}(m_{i},\eta^{\mathrm{FM}}_{k})+ \ldots \; .
\label{eq:free_energy}
\end{equation}

The first term $u^{\mathrm{FM}}_{\mathrm{ani}}(m_{i})$ describes the magnetic anisotropy, depending only on the direction of the magnetization $m_{i}$. The second term $u^{\mathrm{FM}}_{\mathrm{el}}(\eta^{\mathrm{FM}}_{k})$ is a function of the strain components $\eta^{\mathrm{FM}}_{k}$ and thus describes the pure elastic energy density of the ferromagnetic thin film. The dependence of the elastic constants on the magnetization direction $m_{i}$ known as morphic effect is neglected here.\cite{Mueller:58:1940} The third term $u^{\mathrm{FM}}_{\mathrm{magel}}(m_{i},\eta^{\mathrm{FM}}_{k})$ depends on the strain components $\eta^{\mathrm{FM}}_{k}$ and the direction of the magnetization $m_{i}$ and therefore represents the interaction between the elastic and magnetic anisotropy energies,~i.e., the first-order magnetoelastic energy density. Here, the linear coupling between the magnetization and the mechanical strain is left out, since piezomagnetic or even ''pseudo'' piezomagnetic effects are not expected in ferromagnetic materials exhibiting a homogeneous magnetization.\cite{Lacheisserie:book} To account for the shape anisotropy in ferromagnetic thin films with finite dimensions, an additional contribution $u^{\mathrm{FM}}_{\mathrm{demag}}(m_{3})=(\mu_{0}/2)M^2_{\mathrm{s}}m^2_{3}$ is added to Eq.~(\ref{eq:free_energy}).\cite{Morrish:book} Thus, the thermodynamic equilibrium state of the ferromagnetic thin film deposited on top of a ferroelectric substrate is determined by $\tilde{g}^{\mathrm{FM}}$ using the magnetic energy density  $\tilde{u}^{\mathrm{FM}}=u^{\mathrm{FM}}+u^{\mathrm{FM}}_{\mathrm{demag}}(m_{3})$. Since $\tilde{g}^{\mathrm{FM}}$ is independent of the choice of axes, each term has to satisfy the requirements of the crystal symmetry.\cite{Lee:18:1955} In the following, we concentrate on cubic polycrystalline materials, which show no net crystalline magnetic anisotropy. In this case, the remaining energy terms $u^{\mathrm{FM}}_{\mathrm{el}}(\eta_{k}=\eta_{k}^{\mathrm{FM}})$ and $u^{\mathrm{FM}}_{\mathrm{magel}}(m_{i},\eta_{k}=\eta_{k}^{\mathrm{FM}})$ can be expressed as ($i,j\in\{1,2,3\}$):\cite{Lee:18:1955,Birss:book,Sander:62:1999}

\begin{eqnarray}
    u^{\mathrm{FM}}_{\mathrm{el}}&=&\frac{c_{11}}{2}\sum_{i}\eta_{i}^{2}+c_{12}\sum_{i>j}\eta_{i}\eta_{j}+\frac{c_{11}-c_{12}}{4}\sum_{i}\eta_{3+i}^{2} 
    \nonumber \\
    u^{\mathrm{FM}}_{\mathrm{magel}}&=&\chi \bar{B}\left[\sum_{i}\eta_{i}\left(m_{i}^{2}-\frac{1}{3}\right)+\sum_{i>j}\eta_{9-i-j}m_{i}m_{j}\right] \,.
    \label{eq:free_energy_poly}
\end{eqnarray}

Here, $\bar{B}$ denotes the isotropic magnetoelastic coupling coefficient, which is related to the magnetostrictive strain $\bar{\lambda}$ by $\bar{B}=-3(c_{11}-c_{12})\bar{\lambda}/2$.\cite{Lee:18:1955} Furthermore, $c_{kl}$ are the components of the stiffness matrix of the ferromagnetic material with $c_{44}=(c_{11}-c_{12})/2$.\cite{Nye:book} In general, the magnetoelastic coupling coefficient in ferromagnetic thin films deviates from the bulk value due to surface effects and/or the influence of strain.\cite{Sun:66:1991,Tian:79:2009} Since we here use bulk values for $\bar{B}$, a proportionality factor $\chi$ is introduced to account for any deviation in the magnetoelastic coupling from bulk-like behavior. The remaining strain components of the ferromagnetic thin film ($\eta_{k}$ with $k=3,4,5$) can now be calculated to:\cite{Vaz:2008}

\begin{eqnarray}
    \eta_{3}&=&-\frac{c_{12}}{c_{11}}\left({\eta_{1}+\eta_{2}}\right)-\frac{\chi\cdot \bar{B}}{c_{11}}\left(m^{2}_{3}-\frac{1}{3}\right)  
    \nonumber \\
    \eta_{4}&=&-\frac{\chi \bar{B}}{c_{11}-c_{12}}\left(m_{2}m_{3}\right) 
    \nonumber \\
    \eta_{5}&=&-\frac{\chi \bar{B}}{c_{11}-c_{12}}\left(m_{1}m_{3}\right)  \,.
    \label{eq:strainFM}
\end{eqnarray}

To reduce the computational cost, the magnetoelastic terms are neglected in Eq.~(\ref{eq:strainFM}), which yields

\begin{eqnarray}
    \eta_{3}&\approx & -\frac{c_{12}}{c_{11}}\left({\eta_{1}+\eta_{2}}\right)  
    \nonumber \\
    \eta_{4}&\approx & \eta_{5} \approx 0 \,.
    \label{eq:strainFM2}
\end{eqnarray}

This is only valid for in-plane strains $\eta_{1}$ and $\eta_{2}$ larger than $10^{-3}$ using ferromagnetic $3d$ metals like Ni or FeCo. By using the elastic behavior of FeCo published in Refs.~\onlinecite{Hall:31:1960,Clark:103:2008}, the remaining non-zero strain component $\eta_{3}$ of a ferromagnetic FeCo thin film deposited on a miscut BTO crystal as a function of the temperature $T_{\mathrm{MD}}$ can be calculated on the basis of the MD simulations shown in Fig.~\ref{fig:Fig1} with an electric field $E_{\mathrm{z}}=4$\,MV/m applied along the $\mathbf{z}$-direction.

\begin{figure}[tb]
   \includegraphics[width=\columnwidth]{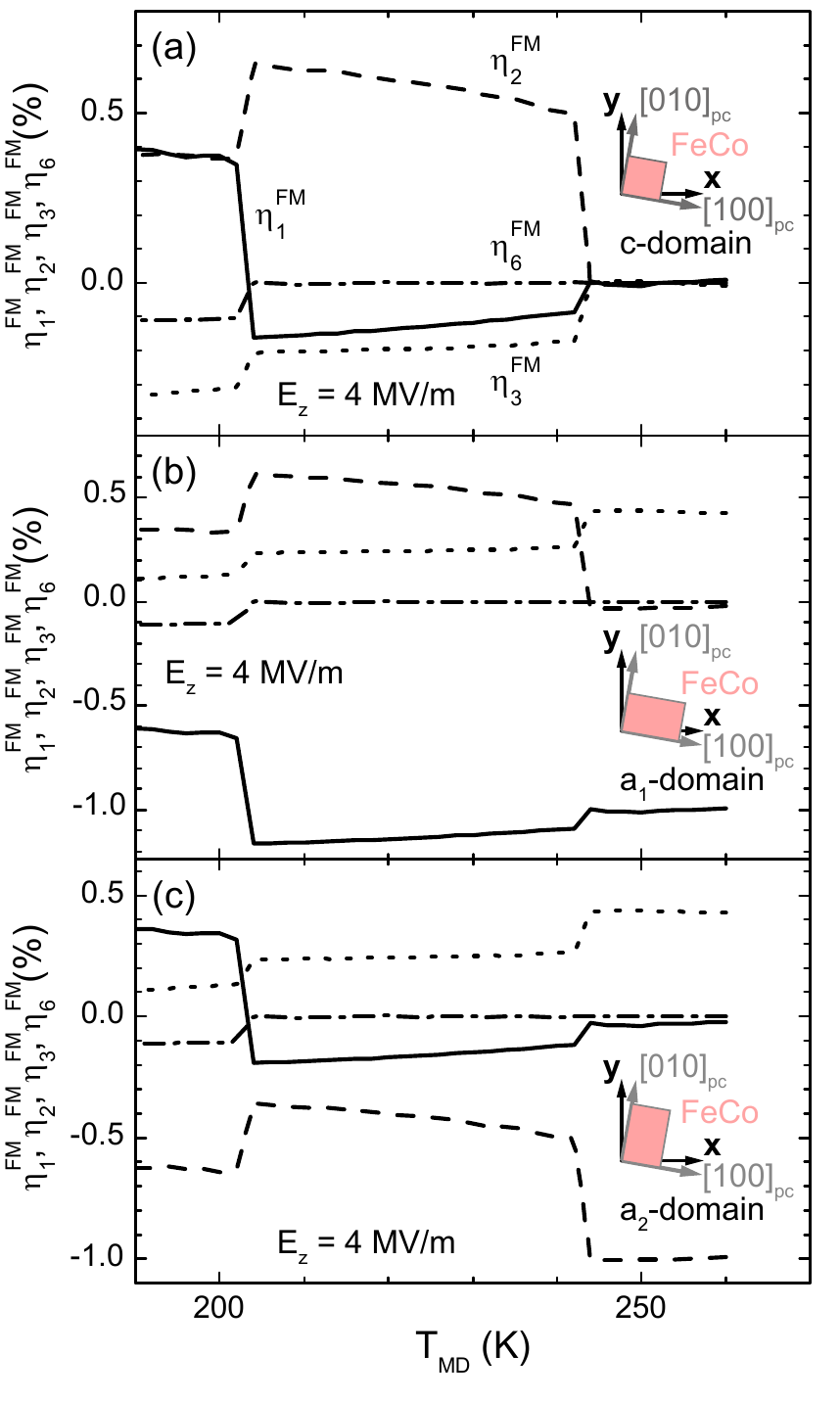}
   \caption{(color online)
Evolution of the non-zero strain components $\eta^{\mathrm{FM}}_{1}$ (solid lines), $\eta^{\mathrm{FM}}_{2}$ (dashed lines), $\eta^{\mathrm{FM}}_{3}$ (dotted lines), and $\eta^{\mathrm{FM}}_{6}$ (dashed-dotted lines) of a FeCo thin film deposited on a miscut BTO crystal as a function of temperature with an electric field of 4\,MV/m applied along the $\mathbf{z}$-direction. The calculations were performed on the basis of the MD simulation shown in Fig.~\ref{fig:Fig1} assuming an ideal elastic coupling between the FeCo thin film and the BTO substrate. The deposition of the unstrained polycrystalline FeCo thin film was assumed to take place at $T_{\mathrm{MD}}=260$\,K on ferroelastic (a) $c$-domains, (b) $a_{1}$-domains, and (c) $a_{2}$-domains.}
    \label{fig:Fig3}
\end{figure}

\begin{figure}[tb]
    \includegraphics[width=\columnwidth]{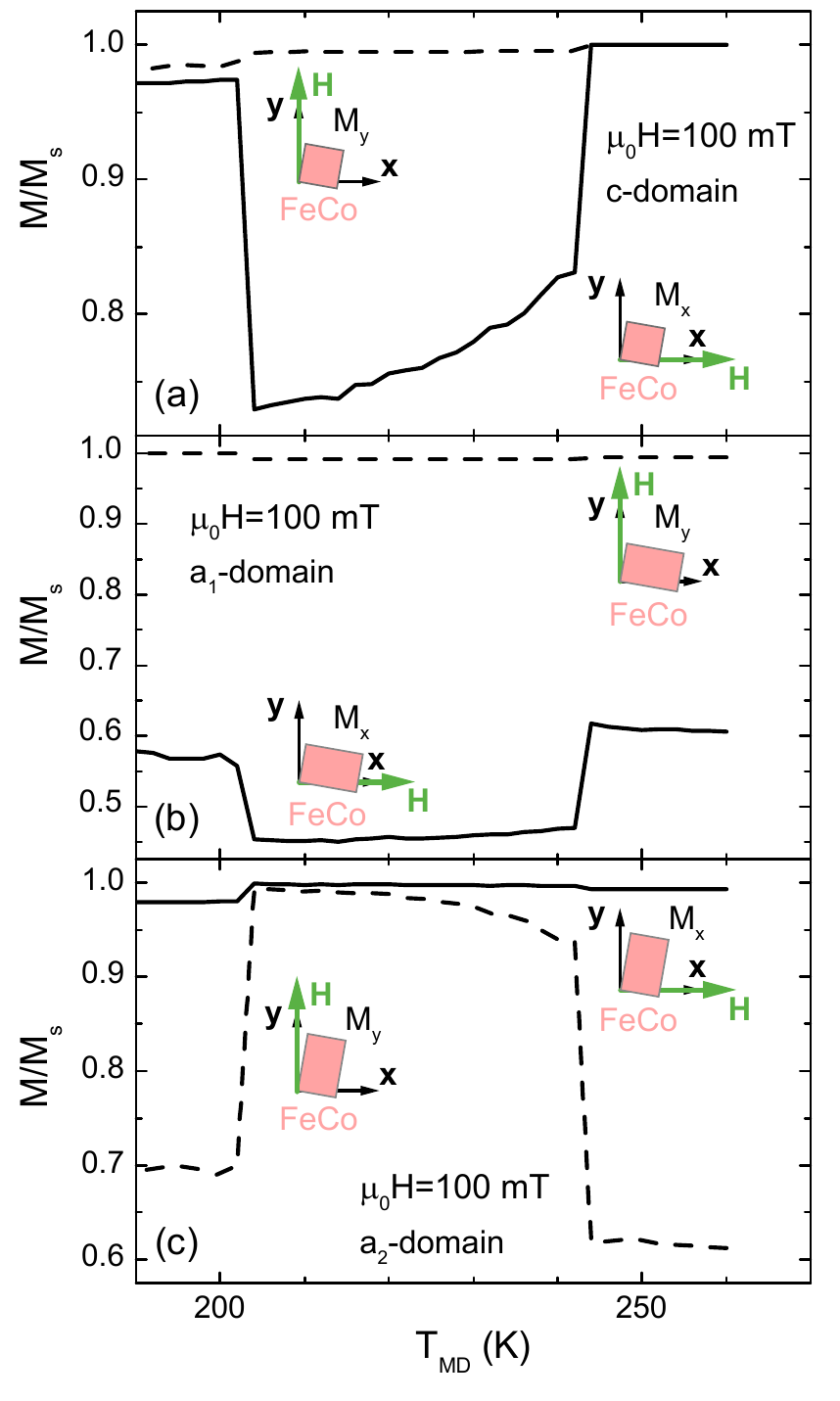}
    \caption{(color online)
Temperature dependence of the normalized magnetization projection $M/M_{\mathrm{s}}$ along an external magnetic field $\mathbf{H}$ with $\mathbf{H}\parallel\mathbf{x}$ ($M_{\mathrm{x}}/M_{\mathrm{s}}$) (solid lines) and $\mathbf{H}\parallel\mathbf{y}$ ($M_{\mathrm{y}}/M_{\mathrm{s}}$) (dashed lines) of a FeCo/BTO multiferroic hybrid. The calculations were performed on the basis of the strain components shown in Fig.~\ref{fig:Fig3} using a magnetic field strength of $\mu_{0}H=100$\,mT and an electric field $E_{\mathrm{z}}=4$\,MV/m. Deposition of the FeCo thin film on ferroelastic (a) $c$-domains, (b) $a_{1}$-domains, and (c) $a_{2}$-domains are considered.}
    \label{fig:Fig4}
\end{figure}

In our experiments, the ferromagnetic polycrystalline thin films are deposited on BTO crystals at room temperature, which corresponds to $T_{\mathrm{MD}}\approx260$\,K [cf.~Fig.~\ref{fig:Fig1}]. At this temperature, the BTO crystal exhibits a certain ferroelastic multi-domain configuration on which the ferromagnetic thin film is deposited in an unstrained state. The detailed domain structure,~i.e., the volume fraction of the tetragonal $c$- and $a$-domains of the BTO substrate during the fabrication process is unknown, since no external fields are applied during the deposition. Thus, the elastic deformation of a FeCo thin film upon cooling is shown in Fig.~\ref{fig:Fig3} assuming a deposition of the ferromagnetic thin film on ferroelastic $c$-domains [Fig.~\ref{fig:Fig3}(a)] as well as on $a_{1}$-domains [Fig.~\ref{fig:Fig3}(b)] and  $a_{2}$-domains [Fig.~\ref{fig:Fig3}(c)]. As expected, large changes of the strain state of the FeCo thin film up to 0.5\% are visible while crossing the ferroelastic phase transitions of the BTO crystal from tetragonal to orthorhombic at $T_{\mathrm{MD}}\approx243$\,K as well as from orthorhombic to rhombohedral at $T_{\mathrm{MD}}\approx203$\,K. These strain changes in turn affect the magnetic properties due to magnetoelastic effects.

\section{Magnetoelastic effects in ferromagnet/BaTiO$\texorpdfstring{_{3}}{3}$ hybrids}
\label{sec:MTtheory}

As the magnetization aligns in such a way that $\tilde{g}^{\mathrm{FM}}$ takes its minimum value, the magnetization direction of the ferromagnetic FeCo thin film in a FeCo/BTO multiferroic hybrid can be calculated as a function of the strain state,~i.e., as a function of the temperature $T_{\mathrm{MD}}$. Using the magnetoelastic behavior of FeCo published in Ref.~\onlinecite{Clark:103:2008} and the correction factor $\chi=1.0$,~i.e., assuming bulk-like magnetoelastic properties, the normalized projection of the magnetization $M/M_{\mathrm{s}}$ along an external magnetic field $\mathbf{H}$ can be calculated.

The normalized magnetization projection with $\mathbf{H}||\mathbf{x}$ ($M_{\mathrm{x}}/M_{\mathrm{s}}$) and $\mathbf{H}||\mathbf{y}$ ($M_{\mathrm{y}}/M_{\mathrm{s}}$) thus obtained is depicted in Fig.~\ref{fig:Fig4}. A magnetic field strength of $\mu_{0}H=100$\,mT was applied for the calculations to ensure a magnetic single domain state. Again, deposition of the FeCo thin film on ferroelastic $c$-domains [Fig.~\ref{fig:Fig4}(a)], $a_{1}$-domains [Fig.~\ref{fig:Fig4}(b)], and $a_{2}$-domains [Fig.~\ref{fig:Fig4}(c)] are considered. As obvious from Fig.~\ref{fig:Fig4}, large modifications of the magnetic state occur at the natural phase transitions of BTO even at a magnetic field strength of $\mu_{0}H=100$\,mT. Assuming the presence of only $c$-domains during the deposition, a strong magnetic anisotropy is expected in the temperature range of the orthorhombic phase of BTO [cf.~Fig.~\ref{fig:Fig4}(a)], while a finite density of $a$-domains mainly causes a huge direction dependence of the magnetization in the tetragonal and rhombohedral phase of BTO [cf.~Fig.~\ref{fig:Fig4}(b) and (c)]. This demonstrates that a different behavior of the magnetization is expected, depending on the volume fraction of the different ferroelastic domains present during deposition. Thus, with the knowledge of the domain configuration during the deposition of the ferromagnetic thin film, predictions of the magnetic behavior of ferromagnetic/BTO multiferroic hybrid structures based on ab-initio calculations are possible.

\section{Experiment versus simulation}
\label{sec:experiment}

To demonstrate the validity of the simulations shown in Fig.~\ref{fig:Fig4}, the magnetic properties of ferromagnet/BTO hybrid structures were investigated experimentally using FeCo and Ni as ferromagnetic materials. These hybrids were fabricated at room temperature by depositing polycrystalline FeCo or Ni thin films with a thickness of 50\,nm onto 0.5\,mm thick (001)-oriented BTO crystals by means of electron beam evaporation at a base pressure of $1.0\times10^{-8}$\,mbar. To prevent oxidation of these thin films, a 10\,nm thick Au film was deposited in situ on top of the ferromagnetic layers. Furthermore, a Au bottom electrode was sputtered on the backside of the BTO crystal. This enables us to apply an electric field across the BTO substrate along the $\mathbf{z}$-direction. After the fabrication process, the ferromagnetic/BTO hybrid structures were heated to 450\,K, well above the ferroelectric transition temperature of BTO, and slowly cooled down to room temperature under an electric field of 400\,kV/m, which results in a preferential formation of ferroelastic $c$-domains in the tetragonal phase of BTO. In the following, we concentrate on two hybrids using either a FeCo or a Ni thin film as the ferromagnetic layer.    

\begin{figure}[tb]
  \includegraphics[width=\columnwidth]{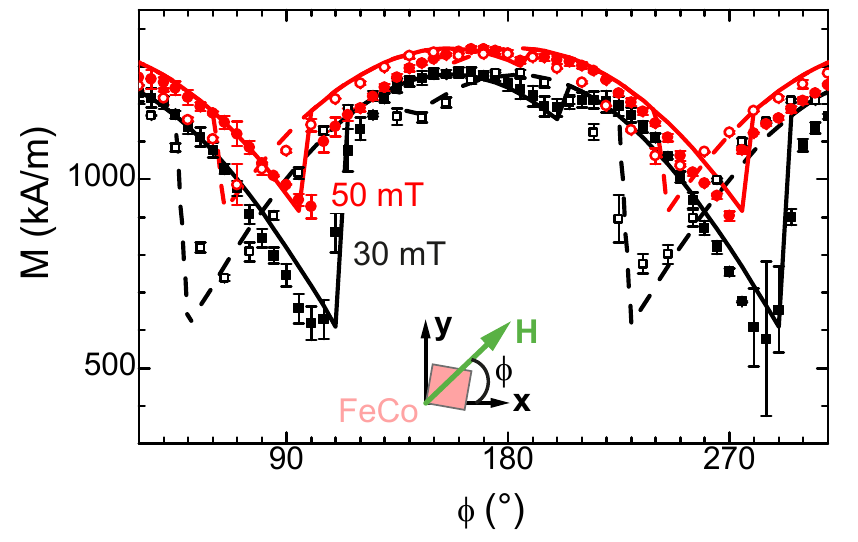}
    \caption{(color online)
SQUID magnetometry measurements of a 50\,nm thick FeCo thin film deposited on a miscut BTO crystal, performed under different in-plane orientations $\phi$ of the external magnetic field. The measurements were carried out for a magnetic field strength of 30\,mT (black symbols) and 50\,mT (red symbols) at 300\,K with an electric field of 400\,kV/m applied along the $\mathbf{z}$-direction of the FeCo/BTO hybrid. After magnetizing the FeCo thin film along the $\mathbf{x}$-direction ($\phi=0^\circ$) with a magnetic field of 1\,T, the angular sweeps were recorded three times in both positive (full symbols) and negative (open symbols) direction of rotation. To simulate the angular dependence of the magnetization by means of magnetoelastic theory (solid and dashed lines), an energy barrier of $\Delta E/M_{\mathrm{s}}=7.5$\,mT was assumed to account for incoherent switching effects of the magnetization. Furthermore, a saturation magnetization of $M_{\mathrm{s}}=1320$\,kA/m was used.}
    \label{fig:Fig5}
\end{figure}

\begin{figure*}[tb]
  \includegraphics[]{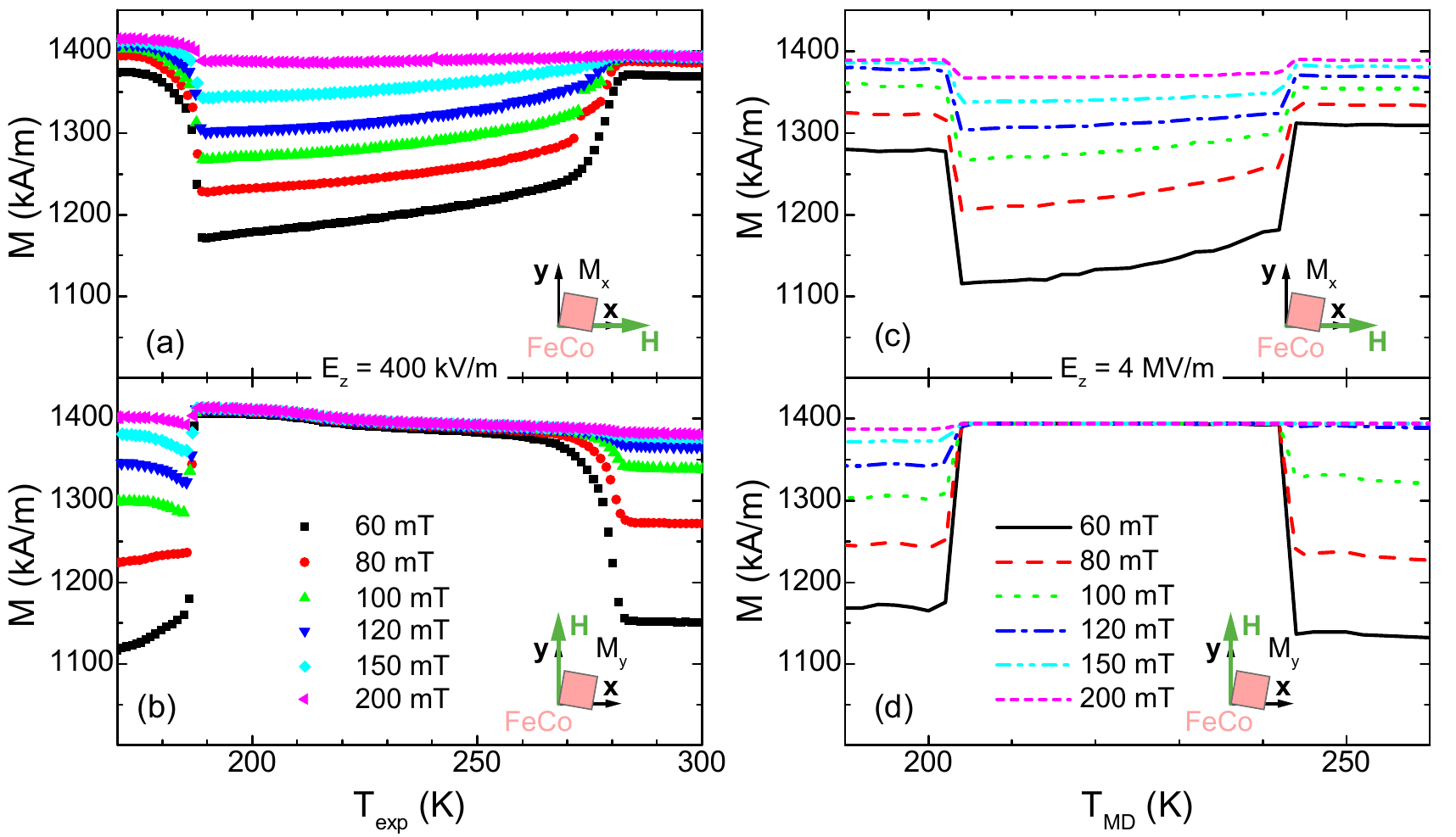}
    \caption{(color online)
[(a), (b)] Temperature dependent magnetization $M(T)$ measured by SQUID magnetometry for different values of the external magnetic field applied along the (a) $\mathbf{x}$- and (b) $\mathbf{y}$-direction of the FeCo/BTO hybrid structure. An electric field of 400\,kV/m was applied along the $\mathbf{z}$-direction during each measurement. [(c), (d)] Simulation of the $M(T)$ behavior assuming the ferroelastic domain sequence as depicted in Fig.~\ref{fig:Fig2}. The volume fraction of the ferroelastic $c$- and $a$-domains during the deposition as well as the proportionality factor $\chi$ was set to $x^{\mathrm{dep}}_{c}=(48\pm5)$\%, $x^{\mathrm{dep}}_{a_{1}}=(12\pm5)$\%, $x^{\mathrm{dep}}_{a_{2}}=(40\pm2)$\%, and $\chi=0.42\pm0.02$ as determined by the angular dependence of the magnetization shown in Fig.~\ref{fig:Fig5}. The temperatures of the experiment and the simulation are denoted by $T_{\mathrm{exp}}$ and $T_{\mathrm{MD}}$, respectively.}
    \label{fig:Fig6}
\end{figure*}

The so far unknown parameters,~i.e.,~the volume fractions of the ferroelastic domains during deposition ($x^{\mathrm{dep}}_{c}$, $x^{\mathrm{dep}}_{a_{1}}$, and $x^{\mathrm{dep}}_{a_{2}}$) and the proportionality factor $\chi$ [cf.~Eq.~(\ref{eq:free_energy_poly})] can be indirectly determined by measuring the magnetic behavior of the ferromagnetic thin film at 300\,K under different in-plane orientations of the external magnetic field while applying an electric field of 400\,kV/m along the $\mathbf{z}$-direction of the BTO substrate,~i.e.,~at a constant strain state of the hybrid structure. As shown in a previous work,\cite{Gepraegs:96:2010} the application of 400\,kV/m ensures a ferroelastic single $c$-domain state in poled BTO crystals at 300\,K. Thus our MD simulations largely overestimate the critical electric field required to obtain a ferroelectric single-domain state. This can be explained by the inhomogeneous nucleation of domains, which are often pinned at defects.\cite{Paul:80:2009} At an electric field strength of 400\,kV/m, parts of the polycrystalline ferromagnetic thin film fabricated on top of ferroelastic $a$-domains are strained, since these $a$-domains are transformed to $c$-domains at $E=400$\,kV/m. Due to magnetoelastic effects, these strained regions of the ferromagnetic thin film cause an angular dependence of the magnetization, which is different for regions of the ferromagnetic thin film deposited on top of $c$-, $a_{1}$- and $a_{2}$-domains. By using SQUID magnetometry, the angular dependence of the magnetization projection $M(\phi)$ along an external magnetic field $\mathbf{H}$ of a ferromagnetic/BTO hybrid structure can be measured.

\begin{figure*}[tb]
  \includegraphics[]{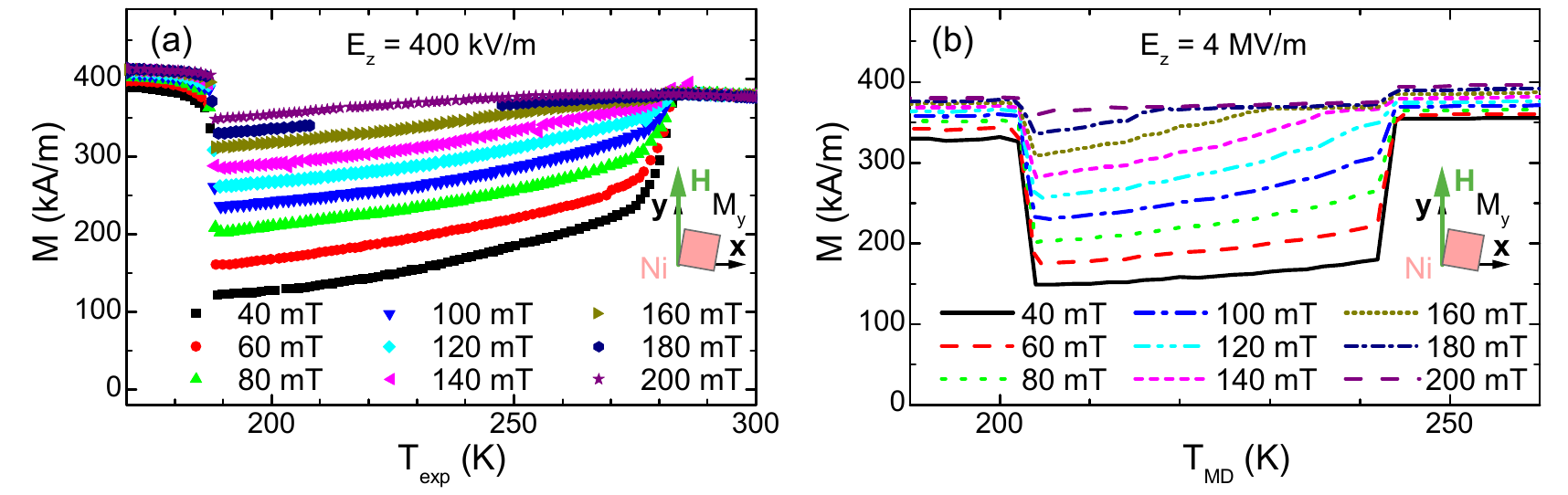}
    \caption{(color online)
Magnetization versus temperature for a Ni/BTO multiferroic hybrid structure. (a) The measurements were carried out with an electric field of $E_{\mathrm{z}}=400$\,kV/m and an external magnetic field $\mathbf{H}\parallel\mathbf{y}$ with different magnetic field strengths. (b) The simulation is carried out using a saturation magnetization of 396\,kA/m, which is determined experimentally by measuring magnetic hysteresis loops. The volume fraction of ferroelastic $c$- and $a$-domains during deposition was set to $x^{\mathrm{dep}}_{a_{1}}=13$\% and $x^{\mathrm{dep}}_{a_{2}}=24$\%. A $\chi$ value of $0.80$ was used.}
    \label{fig:Fig7}
\end{figure*}

As an example, Fig.~\ref{fig:Fig5} shows the angular dependence of $M$ of a FeCo/BTO hybrid structure recorded at external magnetic field strengths of 30\,mT and 50\,mT. Before each measurement, the FeCo thin film was magnetized along the $\mathbf{x}$-direction ($\phi=0^\circ$) using an external magnetic field of 1\,T. The angular sweeps were then recorded three times in both positive and negative directions of rotation. The third back (open symbols)  and forth scans (full symbols) are displayed in Fig.~\ref{fig:Fig5}. The $\phi$-angles at which $M$ exhibits a minimum value mark the direction of a magnetically hard direction. At these positions the measurements carried out on rotating the sample in positive and negative direction do not coincide. The difference as a function of the rotation direction can be attributed to an energy barrier $\Delta E$ describing the energy required to nucleate and unpin domains.\cite{Stone:78:2008} By employing the results of our MD simulations at $T_{\mathrm{MD}}=260$\,K, which corresponds to $T_{\mathrm{exp}}\approx300$\,K, performed under an electric field of 4\,MV/m and 0\,MV/m [cf.~Fig.~\ref{fig:Fig1}], $M(\phi)$ can be calculated using the thermodynamic potential $\tilde{g}^{\mathrm{FM}}$. The energy barrier of $\Delta E/M_{\mathrm{s}}=7.5$\,mT is chosen such that the angles at which the incoherent switching takes place are reproduced by the simulation. Furthermore, the experimentally determined saturation magnetization $M_{\mathrm{s}}=1320$\,kA/m was used. The best fit between experiment and simulation was obtained assuming a concentration of $a$-domains of $x^{\mathrm{dep}}_{a_{1}}=(12\pm5)$\% and $x^{\mathrm{dep}}_{a_{2}}=(40\pm2)$\% during the deposition. We note that the discontinuity observed in the range $130^\circ < \phi < 210^\circ$ gives evidence for the presence of both types of $a$-domains. Furthermore, the proportionality factor $\chi$, was found to be $\chi=0.42\pm0.02$. Using these values, the FeCo/BTO hybrid structure is entirely described in terms of magnetoelastic effects and the temperature dependence of the magnetization can be calculated on the basis of the simulations shown in Fig.~\ref{fig:Fig4}.

The simulation and the experimental results of SQUID magnetometry measurements of the temperature dependent magnetization for different values of the external magnetic field applied along the $\mathbf{x}$- and $\mathbf{y}$-direction of the FeCo/BTO hybrid structure are depicted in Fig.~\ref{fig:Fig6}. The data were recorded while decreasing the temperature from 300\,K to 180\,K after aligning the magnetization into a well-defined state using an external magnetic field of 1\,T. Again, an electric field of 400\,kV/m was applied along the $\mathbf{z}$-direction to ensure the ferroelastic domain sequence in the miscut BTO crystal as shown in Fig.~\ref{fig:Fig2}. In contrast to Sahoo and coworkers,\cite{Sahoo:76:2007} we do not observe any difference in the magnetic behavior upon cooling or heating the hybrid sample. This confirms that the ferroelastic domain control of BTO induces a homogeneous strain state in the ferromagnetic FeCo thin film. Figure~\ref{fig:Fig6} reveals that the simulations of $M$ [cf.~Fig.~\ref{fig:Fig6}(c), (d)] reproduce the experimental results [cf.~Fig.~\ref{fig:Fig6}(a), (b)] fairly well. As discussed in the context of Fig.~\ref{fig:Fig4}, the magnetic anisotropy visible throughout the investigated temperature range can be attributed to the presence of ferroelastic $c$- and $a$-domains during the deposition of FeCo. While the magnetic anisotropy observable in the tetragonal phase of BTO is caused by ferroelastic $a$-domains, the difference of $M_{\mathrm{x}}$ and $M_{\mathrm{y}}$ in the orthorhombic phase can mainly be attributed to a finite volume fraction of ferroelastic $c$-domains. Thus, by controlling the ferroelastic state of the BTO crystal during the fabrication process, the desired magnetic anisotropy can be adjusted in the whole temperature range.

Despite of the large magnetoelastic coupling constants of FeCo, the changes of the magnetization projections $M_{\mathrm{x}}$ and $M_{\mathrm{y}}$ do not exceed 25\%. This is mainly caused by the low $\chi$ value of $0.42\pm0.02$. However, $\chi$ values up to $0.80$ were measured in Ni/BTO hybrid samples. As an example, SQUID magnetometry measurements and the corresponding simulations of the temperature dependent magnetization are shown in Fig.~\ref{fig:Fig7}. The external magnetic field was applied along the $\mathbf{y}$-direction of the hybrid structure. Moreover, the elastic and magnetoelastic constants published in Refs.~\onlinecite{Birss:76:1960,Alers:13:1960} were used for the simulations.
As obvious from Fig.~\ref{fig:Fig7}, an excellent agreement between experiment and simulation was obtained for $x^{\mathrm{dep}}_{a_{1}}=(13\pm5)$\%, $x^{\mathrm{dep}}_{a_{2}}=(24\pm5)$\%, and $\chi=0.80\pm0.05$. In this sample, variations of the magnetization of up to 70\% can be observed at the orthorhombic-rhombohedral phase transition. Unfortunately, these large changes are not persistent and can only be achieved shortly after the deposition. One explanation is the presence of mechanical fatigue, which is expected to lead to two effects. First, mechanical fatigue can occur in the BTO crystal, which in turn reduces the converse piezoelectric strain. Second, the elastic coupling between the ferromagnetic thin film and the BTO crystal might be reduced. In contrast to ferroelectric fatigue effects, the detailed understanding of mechanical fatigue is still lacking. To clarify this issue is of particular technological importance for strain-mediated multiferroic hybrid structures.\cite{Nan:103:2008} However, the excellent agreement between experiment and simulation clearly demonstrates that the magnetoelastic behavior in ferromagnetic/BTO multiferroic hybrid structures is well understood. This paves the way for engineering novel multiferroic composite hybrid structures.

\section{Conclusion}

We have investigated magnetoelastic effects in BTO-based multiferroic hybrid structures using FeCo and Ni as ferromagnetic thin film. As BTO exhibits different ferroelectric/ferroelastic phases, its strain state can be modified by simply controlling the temperature of the BTO crystal. This allowed us to study one and the same ferromagnetic thin film under different elastic constraints. Our detailed analysis showed that a ferroelectric/ferroelastic domain control is mandatory to ensure a predictable, homogeneous strain state of the BTO crystal. To this end, we used miscut BTO substrates and applied a sufficiently large electric field along the out-of-plane direction so that a single ferroelastic domain becomes energetically favorable in each ferroelastic phase. This enabled us to calculate the strain state of the BTO crystal as a function of temperature by means of MD simulations based on a first-principles effective Hamiltonian. Since the ferromagnetic thin film deposited on top of the BTO crystal is elastically clamped to the BTO substrate in our multiferroic hybrid structures, each modification of the in-plane strain components leads to changes of the strain state of the ferromagnetic thin film. Thus, with the knowledge of the elastic behavior of BTO, we could calculate the strain state of the overlying ferromagnetic thin film using a phenomenological thermodynamic model. The calculation revealed, as expected, large strain changes around the natural phase transitions of BTO. Employing magnetoelastic theory, we were able to simulate the magnetic behavior of the ferromagnetic thin film on the basis of the elastic behavior derived from MD simulations. By comparing the results of these simulations to experimental data, we clearly demonstrated that the magnetic properties of BTO-based multiferroic hybrid structures can be theoretically modeled on the basis of first-principles MD simulations. This opens the way to design novel multiferroic composite hybrid structures.

\section*{Acknowledgment}

We thank Thomas Brenninger for continuous technical support and Takeshi Nishimatsu for developing the MD simulation program. Financial support by the German Research Foundation within the priority programs 1157 and 1285 (project Nos. GR~1132/13 \& 14) and the German Excellence Initiative via the \textit{Nanosystems Initiative Munich (NIM)} is gratefully acknowledged.


\begin{thebibliography}{68}


\makeatletter
\providecommand \@ifxundefined [1]{\@ifx{#1\undefined}}
\providecommand \@ifnum [1]{\ifnum #1\expandafter \@firstoftwo \else \expandafter \@secondoftwo \fi}
\providecommand \@ifx [1]{\ifx #1\expandafter \@firstoftwo \else \expandafter \@secondoftwo \fi}
\providecommand \natexlab [1]{#1}\providecommand \enquote  [1]{``#1''}\providecommand \bibnamefont  [1]{#1}
\providecommand \bibfnamefont [1]{#1}
\providecommand \citenamefont [1]{#1}
\providecommand \href@noop [0]{\@secondoftwo}
\providecommand \href [0]{\begingroup \@sanitize@url \@href}
\providecommand \@href[1]{\@@startlink{#1}\@@href}
\providecommand \@@href[1]{\endgroup#1\@@endlink}
\providecommand \@sanitize@url [0]{\catcode `\\12\catcode `\$12\catcode `\&12\catcode `\#12\catcode `\^12\catcode `\_12\catcode `\%12\relax}
\providecommand \@@startlink[1]{}
\providecommand \@@endlink[0]{}
\providecommand \url  [0]{\begingroup\@sanitize@url \@url }
\providecommand \@url [1]{\endgroup\@href {#1}{\urlprefix }}
\providecommand \urlprefix  [0]{URL }
\providecommand \Eprint [0]{\href }
\providecommand \doibase [0]{http://dx.doi.org/}
\providecommand \selectlanguage [0]{\@gobble}
\providecommand \bibinfo  [0]{\@secondoftwo}
\providecommand \bibfield  [0]{\@secondoftwo}
\providecommand \translation [1]{[#1]}
\providecommand \BibitemOpen [0]{}
\providecommand \bibitemStop [0]{}
\providecommand \bibitemNoStop [0]{.\EOS\space}
\providecommand \EOS [0]{\spacefactor3000\relax}
\providecommand \BibitemShut  [1]{\csname bibitem#1\endcsname}
\let\auto@bib@innerbib\@empty



\bibitem [{\citenamefont {Schmid}(1994)}]{Schmid:162:1994}%
  \BibitemOpen
  \bibfield  {author} {\bibinfo {author} {\bibfnamefont {H.}~\bibnamefont
  {Schmid}},\ }\href {\doibase 10.1080/00150199408245120} {\bibfield  {journal}
  {\bibinfo  {journal} {Ferroelectrics}\ }\textbf {\bibinfo {volume} {162}},\
  \bibinfo {pages} {317} (\bibinfo {year} {1994})}\BibitemShut {NoStop}%
\bibitem [{\citenamefont {Eerenstein}\ \emph {et~al.}(2006)\citenamefont
  {Eerenstein}, \citenamefont {Mathur},\ and\ \citenamefont
  {Scott}}]{Eerenstein:442:2006}%
  \BibitemOpen
  \bibfield  {author} {\bibinfo {author} {\bibfnamefont {W.}~\bibnamefont
  {Eerenstein}}, \bibinfo {author} {\bibfnamefont {N.~D.}\ \bibnamefont
  {Mathur}}, \ and\ \bibinfo {author} {\bibfnamefont {J.~F.}\ \bibnamefont
  {Scott}},\ }\href {\doibase 10.1038/nature05023} {\bibfield  {journal}
  {\bibinfo  {journal} {Nature}\ }\textbf {\bibinfo {volume} {442}},\ \bibinfo
  {pages} {759} (\bibinfo {year} {2006})}\BibitemShut {NoStop}%
\bibitem [{\citenamefont {Ramesh}\ and\ \citenamefont
  {Spaldin}(2007)}]{Ramesh:6:2007}%
  \BibitemOpen
  \bibfield  {author} {\bibinfo {author} {\bibfnamefont {R.}~\bibnamefont
  {Ramesh}}\ and\ \bibinfo {author} {\bibfnamefont {N.~A.}\ \bibnamefont
  {Spaldin}},\ }\href {\doibase 10.1038/nmat1805} {\bibfield  {journal}
  {\bibinfo  {journal} {Nat. Mater.}\ }\textbf {\bibinfo {volume} {6}},\
  \bibinfo {pages} {21} (\bibinfo {year} {2007})}\BibitemShut {NoStop}%
\bibitem [{\citenamefont {Spaldin}\ and\ \citenamefont
  {Fiebig}(2005)}]{Spaldin:309:2005}%
  \BibitemOpen
  \bibfield  {author} {\bibinfo {author} {\bibfnamefont {N.~A.}\ \bibnamefont
  {Spaldin}}\ and\ \bibinfo {author} {\bibfnamefont {M.}~\bibnamefont
  {Fiebig}},\ }\href {\doibase 10.1126/science.1113357} {\bibfield  {journal}
  {\bibinfo  {journal} {Science}\ }\textbf {\bibinfo {volume} {309}},\ \bibinfo
  {pages} {391} (\bibinfo {year} {2005})}\BibitemShut {NoStop}%
\bibitem [{\citenamefont {Gepr\"{a}gs}\ \emph {et~al.}(2007)\citenamefont
  {Gepr\"{a}gs}, \citenamefont {Opel}, \citenamefont {Goennenwein},\ and\
  \citenamefont {Gross}}]{Gepraegs:87:2007}%
  \BibitemOpen
  \bibfield  {author} {\bibinfo {author} {\bibfnamefont {S.}~\bibnamefont
  {Gepr\"{a}gs}}, \bibinfo {author} {\bibfnamefont {M.}~\bibnamefont {Opel}},
  \bibinfo {author} {\bibfnamefont {S.~T.~B.}\ \bibnamefont {Goennenwein}}, \
  and\ \bibinfo {author} {\bibfnamefont {R.}~\bibnamefont {Gross}},\ }\href
  {\doibase 10.1080/09500830701194165} {\bibfield  {journal} {\bibinfo
  {journal} {Philos. Mag. Lett.}\ }\textbf {\bibinfo {volume} {87}},\ \bibinfo
  {pages} {141} (\bibinfo {year} {2007})}\BibitemShut {NoStop}%
\bibitem [{\citenamefont {Zheng}\ \emph {et~al.}(2004)\citenamefont {Zheng},
  \citenamefont {Wang}, \citenamefont {Lofland}, \citenamefont {Ma},
  \citenamefont {{Mohaddes-Ardabili}}, \citenamefont {Zhao}, \citenamefont
  {{Salamanca-Riba}}, \citenamefont {Shinde}, \citenamefont {Ogale},
  \citenamefont {Bai}, \citenamefont {Viehland}, \citenamefont {Jia},
  \citenamefont {Schlom}, \citenamefont {Wuttig}, \citenamefont {Roytburd},\
  and\ \citenamefont {Ramesh}}]{Zheng:303:2004}%
  \BibitemOpen
  \bibfield  {author} {\bibinfo {author} {\bibfnamefont {H.}~\bibnamefont
  {Zheng}}, \bibinfo {author} {\bibfnamefont {J.}~\bibnamefont {Wang}},
  \bibinfo {author} {\bibfnamefont {S.~E.}\ \bibnamefont {Lofland}}, \bibinfo
  {author} {\bibfnamefont {Z.}~\bibnamefont {Ma}}, \bibinfo {author}
  {\bibfnamefont {L.}~\bibnamefont {{Mohaddes-Ardabili}}}, \bibinfo {author}
  {\bibfnamefont {T.}~\bibnamefont {Zhao}}, \bibinfo {author} {\bibfnamefont
  {L.}~\bibnamefont {{Salamanca-Riba}}}, \bibinfo {author} {\bibfnamefont
  {S.~R.}\ \bibnamefont {Shinde}}, \bibinfo {author} {\bibfnamefont {S.~B.}\
  \bibnamefont {Ogale}}, \bibinfo {author} {\bibfnamefont {F.}~\bibnamefont
  {Bai}}, \bibinfo {author} {\bibfnamefont {D.}~\bibnamefont {Viehland}},
  \bibinfo {author} {\bibfnamefont {Y.}~\bibnamefont {Jia}}, \bibinfo {author}
  {\bibfnamefont {D.~G.}\ \bibnamefont {Schlom}}, \bibinfo {author}
  {\bibfnamefont {M.}~\bibnamefont {Wuttig}}, \bibinfo {author} {\bibfnamefont
  {A.}~\bibnamefont {Roytburd}}, \ and\ \bibinfo {author} {\bibfnamefont
  {R.}~\bibnamefont {Ramesh}},\ }\href {\doibase 10.1126/science.1094207}
  {\bibfield  {journal} {\bibinfo  {journal} {Science}\ }\textbf {\bibinfo
  {volume} {303}},\ \bibinfo {pages} {661} (\bibinfo {year}
  {2004})}\BibitemShut {NoStop}%
\bibitem [{\citenamefont {Eerenstein}\ \emph {et~al.}(2007)\citenamefont
  {Eerenstein}, \citenamefont {Wiora}, \citenamefont {Prieto}, \citenamefont
  {Scott},\ and\ \citenamefont {Mathur}}]{Eerenstein:6:2007}%
  \BibitemOpen
  \bibfield  {author} {\bibinfo {author} {\bibfnamefont {W.}~\bibnamefont
  {Eerenstein}}, \bibinfo {author} {\bibfnamefont {M.}~\bibnamefont {Wiora}},
  \bibinfo {author} {\bibfnamefont {J.~L.}\ \bibnamefont {Prieto}}, \bibinfo
  {author} {\bibfnamefont {J.~F.}\ \bibnamefont {Scott}}, \ and\ \bibinfo
  {author} {\bibfnamefont {N.~D.}\ \bibnamefont {Mathur}},\ }\href {\doibase
  10.1038/nmat1886} {\bibfield  {journal} {\bibinfo  {journal} {Nat. Mater.}\
  }\textbf {\bibinfo {volume} {6}},\ \bibinfo {pages} {348} (\bibinfo {year}
  {2007})}\BibitemShut {NoStop}%
\bibitem [{\citenamefont {Nan}\ \emph {et~al.}(2008)\citenamefont {Nan},
  \citenamefont {Bichurin}, \citenamefont {Dong}, \citenamefont {Viehland},\
  and\ \citenamefont {Srinivasan}}]{Nan:103:2008}%
  \BibitemOpen
  \bibfield  {author} {\bibinfo {author} {\bibfnamefont {C.-W.}\ \bibnamefont
  {Nan}}, \bibinfo {author} {\bibfnamefont {M.~I.}\ \bibnamefont {Bichurin}},
  \bibinfo {author} {\bibfnamefont {S.}~\bibnamefont {Dong}}, \bibinfo {author}
  {\bibfnamefont {D.}~\bibnamefont {Viehland}}, \ and\ \bibinfo {author}
  {\bibfnamefont {G.}~\bibnamefont {Srinivasan}},\ }\href {\doibase
  10.1063/1.2836410} {\bibfield  {journal} {\bibinfo  {journal} {J. Appl.
  Phys.}\ }\textbf {\bibinfo {volume} {103}},\ \bibinfo {pages} {031101}
  (\bibinfo {year} {2008})}\BibitemShut {NoStop}%
\bibitem [{\citenamefont {Srinivasan}(2010)}]{Srinivasan:40:2010}%
  \BibitemOpen
  \bibfield  {author} {\bibinfo {author} {\bibfnamefont {G.}~\bibnamefont
  {Srinivasan}},\ }\href {\doibase 10.1146/annurev-matsci-070909-104459}
  {\bibfield  {journal} {\bibinfo  {journal} {Ann. Rev. Mater. Res.}\ }\textbf
  {\bibinfo {volume} {40}},\ \bibinfo {pages} {153} (\bibinfo {year}
  {2010})}\BibitemShut {NoStop}%
\bibitem [{\citenamefont {Run}\ \emph {et~al.}(1974)\citenamefont {Run},
  \citenamefont {Terrell},\ and\ \citenamefont {Scholing}}]{Run:9:1974}%
  \BibitemOpen
  \bibfield  {author} {\bibinfo {author} {\bibfnamefont {A.~M. J.~G.}\
  \bibnamefont {Run}}, \bibinfo {author} {\bibfnamefont {D.~R.}\ \bibnamefont
  {Terrell}}, \ and\ \bibinfo {author} {\bibfnamefont {J.~H.}\ \bibnamefont
  {Scholing}},\ }\href {\doibase 10.1007/BF00540771} {\bibfield  {journal}
  {\bibinfo  {journal} {J. Mater. Sci.}\ }\textbf {\bibinfo {volume} {9}},\
  \bibinfo {pages} {1710} (\bibinfo {year} {1974})},\ \bibinfo {note}
  {{10.1007/BF00540771}}\BibitemShut {NoStop}%
\bibitem [{\citenamefont {Ryu}\ \emph {et~al.}(2001)\citenamefont {Ryu},
  \citenamefont {Carazo}, \citenamefont {Uchino},\ and\ \citenamefont
  {Kim}}]{Ryu:40:2001}%
  \BibitemOpen
  \bibfield  {author} {\bibinfo {author} {\bibfnamefont {J.}~\bibnamefont
  {Ryu}}, \bibinfo {author} {\bibfnamefont {A.~V.}\ \bibnamefont {Carazo}},
  \bibinfo {author} {\bibfnamefont {K.}~\bibnamefont {Uchino}}, \ and\ \bibinfo
  {author} {\bibfnamefont {H.-E.}\ \bibnamefont {Kim}},\ }\href {\doibase
  10.1143/JJAP.40.4948} {\bibfield  {journal} {\bibinfo  {journal} {Jpn. J.
  Appl. Phys.}\ }\textbf {\bibinfo {volume} {40}},\ \bibinfo {pages} {4948}
  (\bibinfo {year} {2001})}\BibitemShut {NoStop}%
\bibitem [{\citenamefont {Murugavel}\ \emph {et~al.}(2004)\citenamefont
  {Murugavel}, \citenamefont {Padhan},\ and\ \citenamefont
  {Prellier}}]{Murugavel:85:2004}%
  \BibitemOpen
  \bibfield  {author} {\bibinfo {author} {\bibfnamefont {P.}~\bibnamefont
  {Murugavel}}, \bibinfo {author} {\bibfnamefont {P.}~\bibnamefont {Padhan}}, \
  and\ \bibinfo {author} {\bibfnamefont {W.}~\bibnamefont {Prellier}},\ }\href
  {\doibase 10.1063/1.1825075} {\bibfield  {journal} {\bibinfo  {journal}
  {Appl. Phys. Lett.}\ }\textbf {\bibinfo {volume} {85}},\ \bibinfo {pages}
  {4992} (\bibinfo {year} {2004})}\BibitemShut {NoStop}%
\bibitem [{\citenamefont {Vaz}\ \emph {et~al.}(2010)\citenamefont {Vaz},
  \citenamefont {Hoffman}, \citenamefont {Ahn},\ and\ \citenamefont
  {Ramesh}}]{Vaz:22:2010}%
  \BibitemOpen
  \bibfield  {author} {\bibinfo {author} {\bibfnamefont {C.~A.~F.}\
  \bibnamefont {Vaz}}, \bibinfo {author} {\bibfnamefont {J.}~\bibnamefont
  {Hoffman}}, \bibinfo {author} {\bibfnamefont {C.~H.}\ \bibnamefont {Ahn}}, \
  and\ \bibinfo {author} {\bibfnamefont {R.}~\bibnamefont {Ramesh}},\ }\href
  {\doibase 10.1002/adma.200904326} {\bibfield  {journal} {\bibinfo  {journal}
  {Adv. Mater.}\ }\textbf {\bibinfo {volume} {22}},\ \bibinfo {pages} {2900}
  (\bibinfo {year} {2010})}\BibitemShut {NoStop}%
\bibitem [{\citenamefont {Brandlmaier}\ \emph {et~al.}(2008)\citenamefont
  {Brandlmaier}, \citenamefont {Gepr\"ags}, \citenamefont {Weiler},
  \citenamefont {Boger}, \citenamefont {Opel}, \citenamefont {Huebl},
  \citenamefont {Bihler}, \citenamefont {Brandt}, \citenamefont {Botters},
  \citenamefont {Grundler}, \citenamefont {Gross},\ and\ \citenamefont
  {Goennenwein}}]{Brandlmaier:77:2008}%
  \BibitemOpen
  \bibfield  {author} {\bibinfo {author} {\bibfnamefont {A.}~\bibnamefont
  {Brandlmaier}}, \bibinfo {author} {\bibfnamefont {S.}~\bibnamefont
  {Gepr\"ags}}, \bibinfo {author} {\bibfnamefont {M.}~\bibnamefont {Weiler}},
  \bibinfo {author} {\bibfnamefont {A.}~\bibnamefont {Boger}}, \bibinfo
  {author} {\bibfnamefont {M.}~\bibnamefont {Opel}}, \bibinfo {author}
  {\bibfnamefont {H.}~\bibnamefont {Huebl}}, \bibinfo {author} {\bibfnamefont
  {C.}~\bibnamefont {Bihler}}, \bibinfo {author} {\bibfnamefont {M.~S.}\
  \bibnamefont {Brandt}}, \bibinfo {author} {\bibfnamefont {B.}~\bibnamefont
  {Botters}}, \bibinfo {author} {\bibfnamefont {D.}~\bibnamefont {Grundler}},
  \bibinfo {author} {\bibfnamefont {R.}~\bibnamefont {Gross}}, \ and\ \bibinfo
  {author} {\bibfnamefont {S.~T.~B.}\ \bibnamefont {Goennenwein}},\ }\href
  {\doibase 10.1103/PhysRevB.77.104445} {\bibfield  {journal} {\bibinfo
  {journal} {Phys. Rev. B}\ }\textbf {\bibinfo {volume} {77}},\ \bibinfo
  {pages} {104445} (\bibinfo {year} {2008})}\BibitemShut {NoStop}%
\bibitem [{\citenamefont {Bihler}\ \emph {et~al.}(2008)\citenamefont {Bihler},
  \citenamefont {Althammer}, \citenamefont {Brandlmaier}, \citenamefont
  {Gepr\"ags}, \citenamefont {Weiler}, \citenamefont {Opel}, \citenamefont
  {Schoch}, \citenamefont {Limmer}, \citenamefont {Gross}, \citenamefont
  {Brandt},\ and\ \citenamefont {Goennenwein}}]{Bihler:78:2008}%
  \BibitemOpen
  \bibfield  {author} {\bibinfo {author} {\bibfnamefont {C.}~\bibnamefont
  {Bihler}}, \bibinfo {author} {\bibfnamefont {M.}~\bibnamefont {Althammer}},
  \bibinfo {author} {\bibfnamefont {A.}~\bibnamefont {Brandlmaier}}, \bibinfo
  {author} {\bibfnamefont {S.}~\bibnamefont {Gepr\"ags}}, \bibinfo {author}
  {\bibfnamefont {M.}~\bibnamefont {Weiler}}, \bibinfo {author} {\bibfnamefont
  {M.}~\bibnamefont {Opel}}, \bibinfo {author} {\bibfnamefont {W.}~\bibnamefont
  {Schoch}}, \bibinfo {author} {\bibfnamefont {W.}~\bibnamefont {Limmer}},
  \bibinfo {author} {\bibfnamefont {R.}~\bibnamefont {Gross}}, \bibinfo
  {author} {\bibfnamefont {M.~S.}\ \bibnamefont {Brandt}}, \ and\ \bibinfo
  {author} {\bibfnamefont {S.~T.~B.}\ \bibnamefont {Goennenwein}},\ }\href
  {\doibase 10.1103/PhysRevB.78.045203} {\bibfield  {journal} {\bibinfo
  {journal} {Phys. Rev. B}\ }\textbf {\bibinfo {volume} {78}},\ \bibinfo
  {pages} {045203} (\bibinfo {year} {2008})}\BibitemShut {NoStop}%
\bibitem [{\citenamefont {Weiler}\ \emph {et~al.}(2009)\citenamefont {Weiler},
  \citenamefont {Brandlmaier}, \citenamefont {Gepr\"ags}, \citenamefont
  {Althammer}, \citenamefont {Opel}, \citenamefont {Bihler}, \citenamefont
  {Huebl}, \citenamefont {Brandt}, \citenamefont {Gross},\ and\ \citenamefont
  {Goennenwein}}]{Weiler:11:2009}%
  \BibitemOpen
  \bibfield  {author} {\bibinfo {author} {\bibfnamefont {M.}~\bibnamefont
  {Weiler}}, \bibinfo {author} {\bibfnamefont {A.}~\bibnamefont {Brandlmaier}},
  \bibinfo {author} {\bibfnamefont {S.}~\bibnamefont {Gepr\"ags}}, \bibinfo
  {author} {\bibfnamefont {M.}~\bibnamefont {Althammer}}, \bibinfo {author}
  {\bibfnamefont {M.}~\bibnamefont {Opel}}, \bibinfo {author} {\bibfnamefont
  {C.}~\bibnamefont {Bihler}}, \bibinfo {author} {\bibfnamefont
  {H.}~\bibnamefont {Huebl}}, \bibinfo {author} {\bibfnamefont {M.~S.}\
  \bibnamefont {Brandt}}, \bibinfo {author} {\bibfnamefont {R.}~\bibnamefont
  {Gross}}, \ and\ \bibinfo {author} {\bibfnamefont {S.~T.~B.}\ \bibnamefont
  {Goennenwein}},\ }\href {\doibase 10.1088/1367-2630/11/1/013021} {\bibfield
  {journal} {\bibinfo  {journal} {New J. Phys.}\ }\textbf {\bibinfo {volume}
  {11}},\ \bibinfo {pages} {013021} (\bibinfo {year} {2009})}\BibitemShut
  {NoStop}%
\bibitem [{\citenamefont {Brandlmaier}\ \emph {et~al.}(2011)\citenamefont
  {Brandlmaier}, \citenamefont {Gepr\"{a}gs}, \citenamefont {Woltersdorf},
  \citenamefont {Gross},\ and\ \citenamefont
  {Goennenwein}}]{Brandlmaier:110:2011}%
  \BibitemOpen
  \bibfield  {author} {\bibinfo {author} {\bibfnamefont {A.}~\bibnamefont
  {Brandlmaier}}, \bibinfo {author} {\bibfnamefont {S.}~\bibnamefont
  {Gepr\"{a}gs}}, \bibinfo {author} {\bibfnamefont {G.}~\bibnamefont
  {Woltersdorf}}, \bibinfo {author} {\bibfnamefont {R.}~\bibnamefont {Gross}},
  \ and\ \bibinfo {author} {\bibfnamefont {S.~T.~B.}\ \bibnamefont
  {Goennenwein}},\ }\href {\doibase 10.1063/1.3624663} {\bibfield  {journal}
  {\bibinfo  {journal} {J. Appl. Phys.}\ }\textbf {\bibinfo {volume} {110}},\
  \bibinfo {pages} {043913} (\bibinfo {year} {2011})}\BibitemShut {NoStop}%
\bibitem [{\citenamefont {Israel}\ \emph {et~al.}(2008)\citenamefont {Israel},
  \citenamefont {Mathur},\ and\ \citenamefont {Scott}}]{Israel:7:2008}%
  \BibitemOpen
  \bibfield  {author} {\bibinfo {author} {\bibfnamefont {C.}~\bibnamefont
  {Israel}}, \bibinfo {author} {\bibfnamefont {N.~D.}\ \bibnamefont {Mathur}},
  \ and\ \bibinfo {author} {\bibfnamefont {J.~F.}\ \bibnamefont {Scott}},\
  }\href {\doibase 10.1038/nmat2106} {\bibfield  {journal} {\bibinfo  {journal}
  {Nat. Mater.}\ }\textbf {\bibinfo {volume} {7}},\ \bibinfo {pages} {93}
  (\bibinfo {year} {2008})}\BibitemShut {NoStop}%
\bibitem [{\citenamefont {Ma}\ \emph {et~al.}(2011)\citenamefont {Ma},
  \citenamefont {Hu}, \citenamefont {Li},\ and\ \citenamefont
  {Nan}}]{Ma:23:2011}%
  \BibitemOpen
  \bibfield  {author} {\bibinfo {author} {\bibfnamefont {J.}~\bibnamefont
  {Ma}}, \bibinfo {author} {\bibfnamefont {J.}~\bibnamefont {Hu}}, \bibinfo
  {author} {\bibfnamefont {Z.}~\bibnamefont {Li}}, \ and\ \bibinfo {author}
  {\bibfnamefont {C.-W.}\ \bibnamefont {Nan}},\ }\href {\doibase
  10.1002/adma.201003636} {\bibfield  {journal} {\bibinfo  {journal} {Adv.
  Mater.}\ }\textbf {\bibinfo {volume} {23}},\ \bibinfo {pages} {1062}
  (\bibinfo {year} {2011})}\BibitemShut {NoStop}%
\bibitem [{\citenamefont {van Suchtelen}(1972)}]{Suchtelen:27:1972}%
  \BibitemOpen
  \bibfield  {author} {\bibinfo {author} {\bibfnamefont {J.}~\bibnamefont {van
  Suchtelen}},\ }\href@noop {} {\bibfield  {journal} {\bibinfo  {journal}
  {Philips Res. Rep.}\ }\textbf {\bibinfo {volume} {27}},\ \bibinfo {pages}
  {28} (\bibinfo {year} {1972})}\BibitemShut {NoStop}%
\bibitem [{\citenamefont {Nan}(1994)}]{Nan:50:1994}%
  \BibitemOpen
  \bibfield  {author} {\bibinfo {author} {\bibfnamefont {C.-W.}\ \bibnamefont
  {Nan}},\ }\href {\doibase 10.1103/PhysRevB.50.6082} {\bibfield  {journal}
  {\bibinfo  {journal} {Phys. Rev. B}\ }\textbf {\bibinfo {volume} {50}},\
  \bibinfo {pages} {6082} (\bibinfo {year} {1994})}\BibitemShut {NoStop}%
\bibitem [{\citenamefont {Lee}\ \emph {et~al.}(2000)\citenamefont {Lee},
  \citenamefont {Nath}, \citenamefont {Eom}, \citenamefont {Smoak},\ and\
  \citenamefont {Tsui}}]{Lee:77:2000}%
  \BibitemOpen
  \bibfield  {author} {\bibinfo {author} {\bibfnamefont {M.~K.}\ \bibnamefont
  {Lee}}, \bibinfo {author} {\bibfnamefont {T.~K.}\ \bibnamefont {Nath}},
  \bibinfo {author} {\bibfnamefont {C.~B.}\ \bibnamefont {Eom}}, \bibinfo
  {author} {\bibfnamefont {M.~C.}\ \bibnamefont {Smoak}}, \ and\ \bibinfo
  {author} {\bibfnamefont {F.}~\bibnamefont {Tsui}},\ }\href {\doibase
  10.1063/1.1328762} {\bibfield  {journal} {\bibinfo  {journal} {Appl. Phys.
  Lett.}\ }\textbf {\bibinfo {volume} {77}},\ \bibinfo {pages} {3547} (\bibinfo
  {year} {2000})}\BibitemShut {NoStop}%
\bibitem [{\citenamefont {Dale}\ \emph {et~al.}(2003)\citenamefont {Dale},
  \citenamefont {Fleet}, \citenamefont {Brock},\ and\ \citenamefont
  {Suzuki}}]{Dale:82:2003}%
  \BibitemOpen
  \bibfield  {author} {\bibinfo {author} {\bibfnamefont {D.}~\bibnamefont
  {Dale}}, \bibinfo {author} {\bibfnamefont {A.}~\bibnamefont {Fleet}},
  \bibinfo {author} {\bibfnamefont {J.~D.}\ \bibnamefont {Brock}}, \ and\
  \bibinfo {author} {\bibfnamefont {Y.}~\bibnamefont {Suzuki}},\ }\href
  {\doibase 10.1063/1.1578186} {\bibfield  {journal} {\bibinfo  {journal}
  {Appl. Phys. Lett.}\ }\textbf {\bibinfo {volume} {82}},\ \bibinfo {pages}
  {3725} (\bibinfo {year} {2003})}\BibitemShut {NoStop}%
\bibitem [{\citenamefont {Sahoo}\ \emph {et~al.}(2007)\citenamefont {Sahoo},
  \citenamefont {Polisetty}, \citenamefont {Duan}, \citenamefont {Jaswal},
  \citenamefont {Tsymbal},\ and\ \citenamefont {Binek}}]{Sahoo:76:2007}%
  \BibitemOpen
  \bibfield  {author} {\bibinfo {author} {\bibfnamefont {S.}~\bibnamefont
  {Sahoo}}, \bibinfo {author} {\bibfnamefont {S.}~\bibnamefont {Polisetty}},
  \bibinfo {author} {\bibfnamefont {C.-G.}\ \bibnamefont {Duan}}, \bibinfo
  {author} {\bibfnamefont {S.~S.}\ \bibnamefont {Jaswal}}, \bibinfo {author}
  {\bibfnamefont {E.~Y.}\ \bibnamefont {Tsymbal}}, \ and\ \bibinfo {author}
  {\bibfnamefont {C.}~\bibnamefont {Binek}},\ }\href {\doibase
  10.1103/PhysRevB.76.092108} {\bibfield  {journal} {\bibinfo  {journal} {Phys.
  Rev. B}\ }\textbf {\bibinfo {volume} {76}},\ \bibinfo {pages} {092108}
  (\bibinfo {year} {2007})}\BibitemShut {NoStop}%
\bibitem [{\citenamefont {Taniyama}\ \emph {et~al.}(2009)\citenamefont
  {Taniyama}, \citenamefont {Akasaka}, \citenamefont {Fu},\ and\ \citenamefont
  {Itoh}}]{Taniyama:105:2009}%
  \BibitemOpen
  \bibfield  {author} {\bibinfo {author} {\bibfnamefont {T.}~\bibnamefont
  {Taniyama}}, \bibinfo {author} {\bibfnamefont {K.}~\bibnamefont {Akasaka}},
  \bibinfo {author} {\bibfnamefont {D.}~\bibnamefont {Fu}}, \ and\ \bibinfo
  {author} {\bibfnamefont {M.}~\bibnamefont {Itoh}},\ }\href {\doibase
  10.1063/1.3054357} {\bibfield  {journal} {\bibinfo  {journal} {J. Appl.
  Phys.}\ }\textbf {\bibinfo {volume} {105}},\ \bibinfo {pages} {07D901}
  (\bibinfo {year} {2009})}\BibitemShut {NoStop}%
\bibitem [{\citenamefont {Brivio}\ \emph {et~al.}(2011)\citenamefont {Brivio},
  \citenamefont {Petti}, \citenamefont {Bertacco},\ and\ \citenamefont
  {Cezar}}]{Brivio:98:2011}%
  \BibitemOpen
  \bibfield  {author} {\bibinfo {author} {\bibfnamefont {S.}~\bibnamefont
  {Brivio}}, \bibinfo {author} {\bibfnamefont {D.}~\bibnamefont {Petti}},
  \bibinfo {author} {\bibfnamefont {R.}~\bibnamefont {Bertacco}}, \ and\
  \bibinfo {author} {\bibfnamefont {J.~C.}\ \bibnamefont {Cezar}},\ }\href
  {\doibase 10.1063/1.3554432} {\bibfield  {journal} {\bibinfo  {journal}
  {Appl. Phys. Lett.}\ }\textbf {\bibinfo {volume} {98}},\ \bibinfo {pages}
  {092505} (\bibinfo {year} {2011})}\BibitemShut {NoStop}%
\bibitem [{\citenamefont {Tian}\ \emph {et~al.}(2008)\citenamefont {Tian},
  \citenamefont {Qu}, \citenamefont {Luo}, \citenamefont {Yang}, \citenamefont
  {Guo}, \citenamefont {Zhang}, \citenamefont {Zhao},\ and\ \citenamefont
  {Li}}]{Tian:92:2008}%
  \BibitemOpen
  \bibfield  {author} {\bibinfo {author} {\bibfnamefont {H.~F.}\ \bibnamefont
  {Tian}}, \bibinfo {author} {\bibfnamefont {T.~L.}\ \bibnamefont {Qu}},
  \bibinfo {author} {\bibfnamefont {L.~B.}\ \bibnamefont {Luo}}, \bibinfo
  {author} {\bibfnamefont {J.~J.}\ \bibnamefont {Yang}}, \bibinfo {author}
  {\bibfnamefont {S.~M.}\ \bibnamefont {Guo}}, \bibinfo {author} {\bibfnamefont
  {H.~Y.}\ \bibnamefont {Zhang}}, \bibinfo {author} {\bibfnamefont {Y.~G.}\
  \bibnamefont {Zhao}}, \ and\ \bibinfo {author} {\bibfnamefont {J.~Q.}\
  \bibnamefont {Li}},\ }\href {\doibase 10.1063/1.2844858} {\bibfield
  {journal} {\bibinfo  {journal} {Appl. Phys. Lett.}\ }\textbf {\bibinfo
  {volume} {92}},\ \bibinfo {pages} {063507} (\bibinfo {year}
  {2008})}\BibitemShut {NoStop}%
\bibitem [{\citenamefont {Vaz}\ \emph {et~al.}(2009)\citenamefont {Vaz},
  \citenamefont {Hoffman}, \citenamefont {Posadas},\ and\ \citenamefont
  {Ahn}}]{Vaz:94:2009}%
  \BibitemOpen
  \bibfield  {author} {\bibinfo {author} {\bibfnamefont {C.~A.~F.}\
  \bibnamefont {Vaz}}, \bibinfo {author} {\bibfnamefont {J.}~\bibnamefont
  {Hoffman}}, \bibinfo {author} {\bibfnamefont {A.-B.}\ \bibnamefont
  {Posadas}}, \ and\ \bibinfo {author} {\bibfnamefont {C.~H.}\ \bibnamefont
  {Ahn}},\ }\href {\doibase 10.1063/1.3069280} {\bibfield  {journal} {\bibinfo
  {journal} {Appl. Phys. Lett.}\ }\textbf {\bibinfo {volume} {94}},\ \bibinfo
  {pages} {022504} (\bibinfo {year} {2009})}\BibitemShut {NoStop}%
\bibitem [{\citenamefont {Sterbinsky}\ \emph {et~al.}(2010)\citenamefont
  {Sterbinsky}, \citenamefont {Wessels}, \citenamefont {Kim}, \citenamefont
  {Karapetrova}, \citenamefont {Ryan},\ and\ \citenamefont
  {Keavney}}]{Sterbinsky:96:2010}%
  \BibitemOpen
  \bibfield  {author} {\bibinfo {author} {\bibfnamefont {G.~E.}\ \bibnamefont
  {Sterbinsky}}, \bibinfo {author} {\bibfnamefont {B.~W.}\ \bibnamefont
  {Wessels}}, \bibinfo {author} {\bibfnamefont {J.-W.}\ \bibnamefont {Kim}},
  \bibinfo {author} {\bibfnamefont {E.}~\bibnamefont {Karapetrova}}, \bibinfo
  {author} {\bibfnamefont {P.~J.}\ \bibnamefont {Ryan}}, \ and\ \bibinfo
  {author} {\bibfnamefont {D.~J.}\ \bibnamefont {Keavney}},\ }\href {\doibase
  10.1063/1.3330890} {\bibfield  {journal} {\bibinfo  {journal} {Appl. Phys.
  Lett.}\ }\textbf {\bibinfo {volume} {96}},\ \bibinfo {pages} {092510}
  (\bibinfo {year} {2010})}\BibitemShut {NoStop}%
\bibitem [{\citenamefont {Chopdekar}\ and\ \citenamefont
  {Suzuki}(2006)}]{Chopdekar:89:2006}%
  \BibitemOpen
  \bibfield  {author} {\bibinfo {author} {\bibfnamefont {R.~V.}\ \bibnamefont
  {Chopdekar}}\ and\ \bibinfo {author} {\bibfnamefont {Y.}~\bibnamefont
  {Suzuki}},\ }\href {\doibase 10.1063/1.2370881} {\bibfield  {journal}
  {\bibinfo  {journal} {Appl. Phys. Lett.}\ }\textbf {\bibinfo {volume} {89}},\
  \bibinfo {pages} {182506} (\bibinfo {year} {2006})}\BibitemShut {NoStop}%
\bibitem [{\citenamefont {Czeschka}\ \emph {et~al.}(2009)\citenamefont
  {Czeschka}, \citenamefont {Gepr\"{a}gs}, \citenamefont {Opel}, \citenamefont
  {Goennenwein},\ and\ \citenamefont {Gross}}]{Czeschka:95:2009}%
  \BibitemOpen
  \bibfield  {author} {\bibinfo {author} {\bibfnamefont {F.~D.}\ \bibnamefont
  {Czeschka}}, \bibinfo {author} {\bibfnamefont {S.}~\bibnamefont
  {Gepr\"{a}gs}}, \bibinfo {author} {\bibfnamefont {M.}~\bibnamefont {Opel}},
  \bibinfo {author} {\bibfnamefont {S.~T.~B.}\ \bibnamefont {Goennenwein}}, \
  and\ \bibinfo {author} {\bibfnamefont {R.}~\bibnamefont {Gross}},\ }\href
  {\doibase 10.1063/1.3200236} {\bibfield  {journal} {\bibinfo  {journal}
  {Appl. Phys. Lett.}\ }\textbf {\bibinfo {volume} {95}},\ \bibinfo {pages}
  {062508} (\bibinfo {year} {2009})}\BibitemShut {NoStop}%
\bibitem [{\citenamefont {Gepr\"{a}gs}\ \emph {et~al.}(2009)\citenamefont
  {Gepr\"{a}gs}, \citenamefont {Czeschka}, \citenamefont {Opel}, \citenamefont
  {Goennenwein}, \citenamefont {Yu}, \citenamefont {Mader},\ and\ \citenamefont
  {Gross}}]{Gepraegs:321:2009}%
  \BibitemOpen
  \bibfield  {author} {\bibinfo {author} {\bibfnamefont {S.}~\bibnamefont
  {Gepr\"{a}gs}}, \bibinfo {author} {\bibfnamefont {F.}~\bibnamefont
  {Czeschka}}, \bibinfo {author} {\bibfnamefont {M.}~\bibnamefont {Opel}},
  \bibinfo {author} {\bibfnamefont {S.}~\bibnamefont {Goennenwein}}, \bibinfo
  {author} {\bibfnamefont {W.}~\bibnamefont {Yu}}, \bibinfo {author}
  {\bibfnamefont {W.}~\bibnamefont {Mader}}, \ and\ \bibinfo {author}
  {\bibfnamefont {R.}~\bibnamefont {Gross}},\ }\href {\doibase
  10.1016/j.jmmm.2008.12.029} {\bibfield  {journal} {\bibinfo  {journal} {J.
  Magn. Magn. Mater.}\ }\textbf {\bibinfo {volume} {321}},\ \bibinfo {pages}
  {2001 } (\bibinfo {year} {2009})}\BibitemShut {NoStop}%
\bibitem [{\citenamefont {Opel}\ \emph {et~al.}(2011)\citenamefont {Opel},
  \citenamefont {Gepr\"{a}gs}, \citenamefont {Menzel}, \citenamefont {Nielsen},
  \citenamefont {Reisinger}, \citenamefont {Nielsen}, \citenamefont
  {Brandlmaier}, \citenamefont {Czeschka}, \citenamefont {Althammer},
  \citenamefont {Weiler}, \citenamefont {Goennenwein}, \citenamefont {Simon},
  \citenamefont {Svete}, \citenamefont {Yu}, \citenamefont {H\"{u}hne},
  \citenamefont {Mader},\ and\ \citenamefont {Gross}}]{Opel:208:2011}%
  \BibitemOpen
  \bibfield  {author} {\bibinfo {author} {\bibfnamefont {M.}~\bibnamefont
  {Opel}}, \bibinfo {author} {\bibfnamefont {S.}~\bibnamefont {Gepr\"{a}gs}},
  \bibinfo {author} {\bibfnamefont {E.~P.}\ \bibnamefont {Menzel}}, \bibinfo
  {author} {\bibfnamefont {A.}~\bibnamefont {Nielsen}}, \bibinfo {author}
  {\bibfnamefont {D.}~\bibnamefont {Reisinger}}, \bibinfo {author}
  {\bibfnamefont {K.}~\bibnamefont {Nielsen}}, \bibinfo {author} {\bibfnamefont
  {A.}~\bibnamefont {Brandlmaier}}, \bibinfo {author} {\bibfnamefont {F.~D.}\
  \bibnamefont {Czeschka}}, \bibinfo {author} {\bibfnamefont {M.}~\bibnamefont
  {Althammer}}, \bibinfo {author} {\bibfnamefont {M.}~\bibnamefont {Weiler}},
  \bibinfo {author} {\bibfnamefont {S.~T.~B.}\ \bibnamefont {Goennenwein}},
  \bibinfo {author} {\bibfnamefont {J.}~\bibnamefont {Simon}}, \bibinfo
  {author} {\bibfnamefont {M.}~\bibnamefont {Svete}}, \bibinfo {author}
  {\bibfnamefont {W.}~\bibnamefont {Yu}}, \bibinfo {author} {\bibfnamefont
  {S.}~\bibnamefont {H\"{u}hne}}, \bibinfo {author} {\bibfnamefont
  {W.}~\bibnamefont {Mader}}, \ and\ \bibinfo {author} {\bibfnamefont
  {R.}~\bibnamefont {Gross}},\ }\href {\doibase 10.1002/pssa.201026403}
  {\bibfield  {journal} {\bibinfo  {journal} {Phys. Status Solidi A}\ }\textbf
  {\bibinfo {volume} {208}},\ \bibinfo {pages} {232} (\bibinfo {year}
  {2011})}\BibitemShut {NoStop}%
\bibitem [{\citenamefont {Opel}(2012)}]{Opel:45:2012}%
  \BibitemOpen
  \bibfield  {author} {\bibinfo {author} {\bibfnamefont {M.}~\bibnamefont
  {Opel}},\ }\href {\doibase 10.1088/0022-3727/45/3/033001} {\bibfield
  {journal} {\bibinfo  {journal} {J. Phys. D: Appl. Phys.}\ }\textbf {\bibinfo
  {volume} {45}},\ \bibinfo {pages} {033001} (\bibinfo {year}
  {2012})}\BibitemShut {NoStop}%
\bibitem [{\citenamefont {Sakayori}\ \emph {et~al.}(1995)\citenamefont
  {Sakayori}, \citenamefont {Matsui}, \citenamefont {Abe}, \citenamefont
  {Nakamura}, \citenamefont {Kenmoku}, \citenamefont {Hara}, \citenamefont
  {Ishikawa}, \citenamefont {Kokubu}, \citenamefont {ichi Hirota},\ and\
  \citenamefont {Ikeda}}]{Sakayori:JJAP:34:1995}%
  \BibitemOpen
  \bibfield  {author} {\bibinfo {author} {\bibfnamefont {K.}~\bibnamefont
  {Sakayori}}, \bibinfo {author} {\bibfnamefont {Y.}~\bibnamefont {Matsui}},
  \bibinfo {author} {\bibfnamefont {H.}~\bibnamefont {Abe}}, \bibinfo {author}
  {\bibfnamefont {E.}~\bibnamefont {Nakamura}}, \bibinfo {author}
  {\bibfnamefont {M.}~\bibnamefont {Kenmoku}}, \bibinfo {author} {\bibfnamefont
  {T.}~\bibnamefont {Hara}}, \bibinfo {author} {\bibfnamefont {D.}~\bibnamefont
  {Ishikawa}}, \bibinfo {author} {\bibfnamefont {A.}~\bibnamefont {Kokubu}},
  \bibinfo {author} {\bibfnamefont {K.}~\bibnamefont {ichi Hirota}}, \ and\
  \bibinfo {author} {\bibfnamefont {T.}~\bibnamefont {Ikeda}},\ }\href
  {\doibase 10.1143/JJAP.34.5443} {\bibfield  {journal} {\bibinfo  {journal}
  {Jpn. J. Appl. Phys.}\ }\textbf {\bibinfo {volume} {34}},\ \bibinfo {pages}
  {5443} (\bibinfo {year} {1995})}\BibitemShut {NoStop}%
\bibitem [{\citenamefont {Merz}(1949)}]{Merz:76:1949}%
  \BibitemOpen
  \bibfield  {author} {\bibinfo {author} {\bibfnamefont {W.~J.}\ \bibnamefont
  {Merz}},\ }\href {\doibase 10.1103/PhysRev.76.1221} {\bibfield  {journal}
  {\bibinfo  {journal} {Phys. Rev.}\ }\textbf {\bibinfo {volume} {76}},\
  \bibinfo {pages} {1221} (\bibinfo {year} {1949})}\BibitemShut {NoStop}%
\bibitem [{\citenamefont {Shebanov}(1981)}]{Shebanov:65:1981}%
  \BibitemOpen
  \bibfield  {author} {\bibinfo {author} {\bibfnamefont {L.~A.}\ \bibnamefont
  {Shebanov}},\ }\href {\doibase 10.1002/pssa.2210650137} {\bibfield  {journal}
  {\bibinfo  {journal} {Phys. Status Solidi A}\ }\textbf {\bibinfo {volume}
  {65}},\ \bibinfo {pages} {321} (\bibinfo {year} {1981})}\BibitemShut
  {NoStop}%
\bibitem [{\citenamefont {Nishimatsu}\ \emph {et~al.}(2008)\citenamefont
  {Nishimatsu}, \citenamefont {Waghmare}, \citenamefont {Kawazoe},\ and\
  \citenamefont {Vanderbilt}}]{Nishimatsu:78:2008}%
  \BibitemOpen
  \bibfield  {author} {\bibinfo {author} {\bibfnamefont {T.}~\bibnamefont
  {Nishimatsu}}, \bibinfo {author} {\bibfnamefont {U.~V.}\ \bibnamefont
  {Waghmare}}, \bibinfo {author} {\bibfnamefont {Y.}~\bibnamefont {Kawazoe}}, \
  and\ \bibinfo {author} {\bibfnamefont {D.}~\bibnamefont {Vanderbilt}},\
  }\href {\doibase 10.1103/PhysRevB.78.104104} {\bibfield  {journal} {\bibinfo
  {journal} {Phys. Rev. B}\ }\textbf {\bibinfo {volume} {78}},\ \bibinfo
  {pages} {104104} (\bibinfo {year} {2008})}\BibitemShut {NoStop}%
\bibitem [{\citenamefont {Nishimatsu}()}]{FERAMcode}%
  \BibitemOpen
  \bibfield  {author} {\bibinfo {author} {\bibfnamefont {T.}~\bibnamefont
  {Nishimatsu}},\ }\href@noop {} {\enquote {\bibinfo {title} {Feram~code},}\
  }\bibinfo {howpublished} {http://loto.sourceforge.net/feram/}\BibitemShut
  {NoStop}%
\bibitem [{\citenamefont {Nishimatsu}\ \emph {et~al.}(2010)\citenamefont
  {Nishimatsu}, \citenamefont {Iwamoto}, \citenamefont {Kawazoe},\ and\
  \citenamefont {Waghmare}}]{Nishimatsu:82:2010}%
  \BibitemOpen
  \bibfield  {author} {\bibinfo {author} {\bibfnamefont {T.}~\bibnamefont
  {Nishimatsu}}, \bibinfo {author} {\bibfnamefont {M.}~\bibnamefont {Iwamoto}},
  \bibinfo {author} {\bibfnamefont {Y.}~\bibnamefont {Kawazoe}}, \ and\
  \bibinfo {author} {\bibfnamefont {U.~V.}\ \bibnamefont {Waghmare}},\ }\href
  {\doibase 10.1103/PhysRevB.82.134106} {\bibfield  {journal} {\bibinfo
  {journal} {Phys. Rev. B}\ }\textbf {\bibinfo {volume} {82}},\ \bibinfo
  {pages} {134106} (\bibinfo {year} {2010})}\BibitemShut {NoStop}%
\bibitem [{\citenamefont {Paul}\ \emph {et~al.}(2007)\citenamefont {Paul},
  \citenamefont {Nishimatsu}, \citenamefont {Kawazoe},\ and\ \citenamefont
  {Waghmare}}]{Paul:99:2007}%
  \BibitemOpen
  \bibfield  {author} {\bibinfo {author} {\bibfnamefont {J.}~\bibnamefont
  {Paul}}, \bibinfo {author} {\bibfnamefont {T.}~\bibnamefont {Nishimatsu}},
  \bibinfo {author} {\bibfnamefont {Y.}~\bibnamefont {Kawazoe}}, \ and\
  \bibinfo {author} {\bibfnamefont {U.~V.}\ \bibnamefont {Waghmare}},\ }\href
  {\doibase 10.1103/PhysRevLett.99.077601} {\bibfield  {journal} {\bibinfo
  {journal} {Phys. Rev. Lett.}\ }\textbf {\bibinfo {volume} {99}},\ \bibinfo
  {pages} {077601} (\bibinfo {year} {2007})}\BibitemShut {NoStop}%
\bibitem [{\citenamefont {Garc\'{\i}a}\ and\ \citenamefont
  {Vanderbilt}(1998)}]{Garcia:72:2981}%
  \BibitemOpen
  \bibfield  {author} {\bibinfo {author} {\bibfnamefont {A.}~\bibnamefont
  {Garc\'{\i}a}}\ and\ \bibinfo {author} {\bibfnamefont {D.}~\bibnamefont
  {Vanderbilt}},\ }\href {\doibase 10.1063/1.121514} {\bibfield  {journal}
  {\bibinfo  {journal} {Appl. Phys. Lett.}\ }\textbf {\bibinfo {volume} {72}},\
  \bibinfo {pages} {2981} (\bibinfo {year} {1998})}\BibitemShut {NoStop}%
\bibitem [{\citenamefont {Catalan}\ \emph {et~al.}(2012)\citenamefont
  {Catalan}, \citenamefont {Seidel}, \citenamefont {Ramesh},\ and\
  \citenamefont {Scott}}]{Catalan:84:2012}%
  \BibitemOpen
  \bibfield  {author} {\bibinfo {author} {\bibfnamefont {G.}~\bibnamefont
  {Catalan}}, \bibinfo {author} {\bibfnamefont {J.}~\bibnamefont {Seidel}},
  \bibinfo {author} {\bibfnamefont {R.}~\bibnamefont {Ramesh}}, \ and\ \bibinfo
  {author} {\bibfnamefont {J.~F.}\ \bibnamefont {Scott}},\ }\href {\doibase
  10.1103/RevModPhys.84.119} {\bibfield  {journal} {\bibinfo  {journal} {Rev.
  Mod. Phys.}\ }\textbf {\bibinfo {volume} {84}},\ \bibinfo {pages} {119}
  (\bibinfo {year} {2012})}\BibitemShut {NoStop}%
\bibitem [{\citenamefont {Damjanovic}(1998)}]{Damjanovic:61:1998}%
  \BibitemOpen
  \bibfield  {author} {\bibinfo {author} {\bibfnamefont {D.}~\bibnamefont
  {Damjanovic}},\ }\href {\doibase 10.1088/0034-4885/61/9/002} {\bibfield
  {journal} {\bibinfo  {journal} {Rep. Prog. Phys.}\ }\textbf {\bibinfo
  {volume} {61}},\ \bibinfo {pages} {1267} (\bibinfo {year}
  {1998})}\BibitemShut {NoStop}%
\bibitem [{\citenamefont {Park}\ \emph {et~al.}(1999)\citenamefont {Park},
  \citenamefont {Wada}, \citenamefont {Cross},\ and\ \citenamefont
  {Shrout}}]{Park:86:1999}%
  \BibitemOpen
  \bibfield  {author} {\bibinfo {author} {\bibfnamefont {S.-E.}\ \bibnamefont
  {Park}}, \bibinfo {author} {\bibfnamefont {S.}~\bibnamefont {Wada}}, \bibinfo
  {author} {\bibfnamefont {L.~E.}\ \bibnamefont {Cross}}, \ and\ \bibinfo
  {author} {\bibfnamefont {T.~R.}\ \bibnamefont {Shrout}},\ }\href {\doibase
  10.1063/1.371120} {\bibfield  {journal} {\bibinfo  {journal} {J. Appl.
  Phys.}\ }\textbf {\bibinfo {volume} {86}},\ \bibinfo {pages} {2746} (\bibinfo
  {year} {1999})}\BibitemShut {NoStop}%
\bibitem [{\citenamefont {Fu}\ and\ \citenamefont {Cohen}(2000)}]{Fu:403:2000}%
  \BibitemOpen
  \bibfield  {author} {\bibinfo {author} {\bibfnamefont {H.}~\bibnamefont
  {Fu}}\ and\ \bibinfo {author} {\bibfnamefont {R.~E.}\ \bibnamefont {Cohen}},\
  }\href {\doibase 10.1038/35002022} {\bibfield  {journal} {\bibinfo  {journal}
  {Nature}\ }\textbf {\bibinfo {volume} {403}},\ \bibinfo {pages} {281}
  (\bibinfo {year} {2000})}\BibitemShut {NoStop}%
\bibitem [{\citenamefont {Vanderbilt}\ and\ \citenamefont
  {Cohen}(2001)}]{Vanderbilt:63:2001}%
  \BibitemOpen
  \bibfield  {author} {\bibinfo {author} {\bibfnamefont {D.}~\bibnamefont
  {Vanderbilt}}\ and\ \bibinfo {author} {\bibfnamefont {M.~H.}\ \bibnamefont
  {Cohen}},\ }\href {\doibase 10.1103/PhysRevB.63.094108} {\bibfield  {journal}
  {\bibinfo  {journal} {Phys. Rev. B}\ }\textbf {\bibinfo {volume} {63}},\
  \bibinfo {pages} {094108} (\bibinfo {year} {2001})}\BibitemShut {NoStop}%
\bibitem [{\citenamefont {Paul}\ \emph {et~al.}(2009)\citenamefont {Paul},
  \citenamefont {Nishimatsu}, \citenamefont {Kawazoe},\ and\ \citenamefont
  {Waghmare}}]{Paul:80:2009}%
  \BibitemOpen
  \bibfield  {author} {\bibinfo {author} {\bibfnamefont {J.}~\bibnamefont
  {Paul}}, \bibinfo {author} {\bibfnamefont {T.}~\bibnamefont {Nishimatsu}},
  \bibinfo {author} {\bibfnamefont {Y.}~\bibnamefont {Kawazoe}}, \ and\
  \bibinfo {author} {\bibfnamefont {U.~V.}\ \bibnamefont {Waghmare}},\ }\href
  {\doibase 10.1103/PhysRevB.80.024107} {\bibfield  {journal} {\bibinfo
  {journal} {Phys. Rev. B}\ }\textbf {\bibinfo {volume} {80}},\ \bibinfo
  {pages} {024107} (\bibinfo {year} {2009})}\BibitemShut {NoStop}%
\bibitem [{\citenamefont {Marcus}\ and\ \citenamefont
  {Jona}(1994)}]{Marcus:55:1994}%
  \BibitemOpen
  \bibfield  {author} {\bibinfo {author} {\bibfnamefont {P.}~\bibnamefont
  {Marcus}}\ and\ \bibinfo {author} {\bibfnamefont {F.}~\bibnamefont {Jona}},\
  }\href {\doibase 10.1016/0022-3697(94)90577-0} {\bibfield  {journal}
  {\bibinfo  {journal} {J. Phys. Chem. Solids}\ }\textbf {\bibinfo {volume}
  {55}},\ \bibinfo {pages} {1513 } (\bibinfo {year} {1994})}\BibitemShut
  {NoStop}%
\bibitem [{\citenamefont {Pertsev}\ \emph {et~al.}(1998)\citenamefont
  {Pertsev}, \citenamefont {Zembilgotov},\ and\ \citenamefont
  {Tagantsev}}]{Pertsev:80:1998}%
  \BibitemOpen
  \bibfield  {author} {\bibinfo {author} {\bibfnamefont {N.~A.}\ \bibnamefont
  {Pertsev}}, \bibinfo {author} {\bibfnamefont {A.~G.}\ \bibnamefont
  {Zembilgotov}}, \ and\ \bibinfo {author} {\bibfnamefont {A.~K.}\ \bibnamefont
  {Tagantsev}},\ }\href {\doibase 10.1103/PhysRevLett.80.1988} {\bibfield
  {journal} {\bibinfo  {journal} {Phys. Rev. Lett.}\ }\textbf {\bibinfo
  {volume} {80}},\ \bibinfo {pages} {1988} (\bibinfo {year}
  {1998})}\BibitemShut {NoStop}%
\bibitem [{\citenamefont {Stoner}\ and\ \citenamefont
  {Wohlfarth}(1948)}]{Stoner:240:1948}%
  \BibitemOpen
  \bibfield  {author} {\bibinfo {author} {\bibfnamefont {E.~C.}\ \bibnamefont
  {Stoner}}\ and\ \bibinfo {author} {\bibfnamefont {E.~P.}\ \bibnamefont
  {Wohlfarth}},\ }\href {\doibase 10.1098/rsta.1948.0007} {\bibfield  {journal}
  {\bibinfo  {journal} {Philos. Trans. R. Soc. London, Ser. A}\ }\textbf
  {\bibinfo {volume} {240}},\ \bibinfo {pages} {599} (\bibinfo {year}
  {1948})}\BibitemShut {NoStop}%
\bibitem [{\citenamefont {{O'Handley}}(2000)}]{OHandley:book}%
  \BibitemOpen
  \bibfield  {author} {\bibinfo {author} {\bibfnamefont {R.~C.}\ \bibnamefont
  {{O'Handley}}},\ }\href@noop {} {\emph {\bibinfo {title} {Modern Magnetic
  Materials: Principles and Applications}}},\ \bibinfo {edition} {1st}\ ed.\
  (\bibinfo  {publisher} {John Wiley \& Sons},\ \bibinfo {year}
  {2000})\BibitemShut {NoStop}%
\bibitem [{\citenamefont {Mueller}(1940)}]{Mueller:58:1940}%
  \BibitemOpen
  \bibfield  {author} {\bibinfo {author} {\bibfnamefont {H.}~\bibnamefont
  {Mueller}},\ }\href {\doibase 10.1103/PhysRev.58.805} {\bibfield  {journal}
  {\bibinfo  {journal} {Phys. Rev.}\ }\textbf {\bibinfo {volume} {58}},\
  \bibinfo {pages} {805} (\bibinfo {year} {1940})}\BibitemShut {NoStop}%
\bibitem [{\citenamefont {du~Tr\'{e}molet~de
  Lacheisserie}(1993)}]{Lacheisserie:book}%
  \BibitemOpen
  \bibfield  {author} {\bibinfo {author} {\bibfnamefont {E.}~\bibnamefont
  {du~Tr\'{e}molet~de Lacheisserie}},\ }\href@noop {} {\emph {\bibinfo {title}
  {Magnetostriction: Theory and Applications of Magnetoelasticity}}}\ (\bibinfo
   {publisher} {CRC Press},\ \bibinfo {year} {1993})\BibitemShut {NoStop}%
\bibitem [{\citenamefont {Morrish}(2001)}]{Morrish:book}%
  \BibitemOpen
  \bibfield  {author} {\bibinfo {author} {\bibfnamefont {A.~H.}\ \bibnamefont
  {Morrish}},\ }\href@noop {} {\emph {\bibinfo {title} {The Physical Principles
  of Magnetism}}},\ \bibinfo {edition} {1st}\ ed.\ (\bibinfo  {publisher}
  {{Wiley-IEEE} Press},\ \bibinfo {year} {2001})\BibitemShut {NoStop}%
\bibitem [{\citenamefont {Lee}(1955)}]{Lee:18:1955}%
  \BibitemOpen
  \bibfield  {author} {\bibinfo {author} {\bibfnamefont {E.~W.}\ \bibnamefont
  {Lee}},\ }\href {\doibase 10.1088/0034-4885/18/1/305} {\bibfield  {journal}
  {\bibinfo  {journal} {Rep. Prog. Phys.}\ }\textbf {\bibinfo {volume} {18}},\
  \bibinfo {pages} {184} (\bibinfo {year} {1955})}\BibitemShut {NoStop}%
\bibitem [{\citenamefont {Birss}(1966)}]{Birss:book}%
  \BibitemOpen
  \bibfield  {author} {\bibinfo {author} {\bibfnamefont {R.~R.}\ \bibnamefont
  {Birss}},\ }\href@noop {} {\emph {\bibinfo {title} {Symmetry and
  magnetism}}}\ (\bibinfo  {publisher} {{North-Holland} Pub. Co.},\ \bibinfo
  {year} {1966})\BibitemShut {NoStop}%
\bibitem [{\citenamefont {Sander}(1999)}]{Sander:62:1999}%
  \BibitemOpen
  \bibfield  {author} {\bibinfo {author} {\bibfnamefont {D.}~\bibnamefont
  {Sander}},\ }\href {\doibase 10.1088/0034-4885/62/5/204} {\bibfield
  {journal} {\bibinfo  {journal} {Rep. Prog. Phys.}\ }\textbf {\bibinfo
  {volume} {62}},\ \bibinfo {pages} {809} (\bibinfo {year} {1999})}\BibitemShut
  {NoStop}%
\bibitem [{\citenamefont {Nye}(1985)}]{Nye:book}%
  \BibitemOpen
  \bibfield  {author} {\bibinfo {author} {\bibfnamefont {J.}~\bibnamefont
  {Nye}},\ }\href@noop {} {\emph {\bibinfo {title} {Physical Properties of
  Crystals}}}\ (\bibinfo  {publisher} {Oxford University Press},\ \bibinfo
  {year} {1985})\BibitemShut {NoStop}%
\bibitem [{\citenamefont {Sun}\ and\ \citenamefont
  {O'Handley}(1991)}]{Sun:66:1991}%
  \BibitemOpen
  \bibfield  {author} {\bibinfo {author} {\bibfnamefont {S.~W.}\ \bibnamefont
  {Sun}}\ and\ \bibinfo {author} {\bibfnamefont {R.~C.}\ \bibnamefont
  {O'Handley}},\ }\href {\doibase 10.1103/PhysRevLett.66.2798} {\bibfield
  {journal} {\bibinfo  {journal} {Phys. Rev. Lett.}\ }\textbf {\bibinfo
  {volume} {66}},\ \bibinfo {pages} {2798} (\bibinfo {year}
  {1991})}\BibitemShut {NoStop}%
\bibitem [{\citenamefont {Tian}\ \emph {et~al.}(2009)\citenamefont {Tian},
  \citenamefont {Sander},\ and\ \citenamefont {Kirschner}}]{Tian:79:2009}%
  \BibitemOpen
  \bibfield  {author} {\bibinfo {author} {\bibfnamefont {Z.}~\bibnamefont
  {Tian}}, \bibinfo {author} {\bibfnamefont {D.}~\bibnamefont {Sander}}, \ and\
  \bibinfo {author} {\bibfnamefont {J.}~\bibnamefont {Kirschner}},\ }\href
  {\doibase 10.1103/PhysRevB.79.024432} {\bibfield  {journal} {\bibinfo
  {journal} {Phys. Rev. B}\ }\textbf {\bibinfo {volume} {79}},\ \bibinfo
  {pages} {024432} (\bibinfo {year} {2009})}\BibitemShut {NoStop}%
\bibitem [{\citenamefont {Vaz}(2008)}]{Vaz:2008}%
  \BibitemOpen
  \bibfield  {author} {\bibinfo {author} {\bibfnamefont {C.~A.~F.}\
  \bibnamefont {Vaz}},\ }\href {http://arxiv.org/abs/0811.2146} {\bibfield
  {journal} {\bibinfo  {journal} {{arXiv:0811.2146}}\ } (\bibinfo {year}
  {2008})}\BibitemShut {NoStop}%
\bibitem [{\citenamefont {Hall}(1960)}]{Hall:31:1960}%
  \BibitemOpen
  \bibfield  {author} {\bibinfo {author} {\bibfnamefont {R.~C.}\ \bibnamefont
  {Hall}},\ }\href {\doibase 10.1063/1.1984643} {\bibfield  {journal} {\bibinfo
   {journal} {J. Appl. Phys.}\ }\textbf {\bibinfo {volume} {31}},\ \bibinfo
  {pages} {S157} (\bibinfo {year} {1960})}\BibitemShut {NoStop}%
\bibitem [{\citenamefont {Clark}\ \emph {et~al.}(2008)\citenamefont {Clark},
  \citenamefont {Restorff}, \citenamefont {WunFogle}, \citenamefont {Wu},\ and\
  \citenamefont {Lograsso}}]{Clark:103:2008}%
  \BibitemOpen
  \bibfield  {author} {\bibinfo {author} {\bibfnamefont {A.~E.}\ \bibnamefont
  {Clark}}, \bibinfo {author} {\bibfnamefont {J.~B.}\ \bibnamefont {Restorff}},
  \bibinfo {author} {\bibfnamefont {M.}~\bibnamefont {WunFogle}}, \bibinfo
  {author} {\bibfnamefont {D.}~\bibnamefont {Wu}}, \ and\ \bibinfo {author}
  {\bibfnamefont {T.~A.}\ \bibnamefont {Lograsso}},\ }\href {\doibase
  10.1063/1.2831360} {\bibfield  {journal} {\bibinfo  {journal} {J. Appl.
  Phys.}\ }\textbf {\bibinfo {volume} {103}},\ \bibinfo {pages} {07B310}
  (\bibinfo {year} {2008})}\BibitemShut {NoStop}%
\bibitem [{\citenamefont {Gepr\"{a}gs}\ \emph {et~al.}(2010)\citenamefont
  {Gepr\"{a}gs}, \citenamefont {Brandlmaier}, \citenamefont {Opel},
  \citenamefont {Gross},\ and\ \citenamefont {Goennenwein}}]{Gepraegs:96:2010}%
  \BibitemOpen
  \bibfield  {author} {\bibinfo {author} {\bibfnamefont {S.}~\bibnamefont
  {Gepr\"{a}gs}}, \bibinfo {author} {\bibfnamefont {A.}~\bibnamefont
  {Brandlmaier}}, \bibinfo {author} {\bibfnamefont {M.}~\bibnamefont {Opel}},
  \bibinfo {author} {\bibfnamefont {R.}~\bibnamefont {Gross}}, \ and\ \bibinfo
  {author} {\bibfnamefont {S.~T.~B.}\ \bibnamefont {Goennenwein}},\ }\href
  {\doibase 10.1063/1.3377923} {\bibfield  {journal} {\bibinfo  {journal}
  {Appl. Phys. Lett.}\ }\textbf {\bibinfo {volume} {96}},\ \bibinfo {pages}
  {142509} (\bibinfo {year} {2010})}\BibitemShut {NoStop}%
\bibitem [{\citenamefont {Stone}\ \emph {et~al.}(2008)\citenamefont {Stone},
  \citenamefont {Bihler}, \citenamefont {Kraus}, \citenamefont {Scarpulla},
  \citenamefont {Beeman}, \citenamefont {Yu}, \citenamefont {Brandt},\ and\
  \citenamefont {Dubon}}]{Stone:78:2008}%
  \BibitemOpen
  \bibfield  {author} {\bibinfo {author} {\bibfnamefont {P.~R.}\ \bibnamefont
  {Stone}}, \bibinfo {author} {\bibfnamefont {C.}~\bibnamefont {Bihler}},
  \bibinfo {author} {\bibfnamefont {M.}~\bibnamefont {Kraus}}, \bibinfo
  {author} {\bibfnamefont {M.~A.}\ \bibnamefont {Scarpulla}}, \bibinfo {author}
  {\bibfnamefont {J.~W.}\ \bibnamefont {Beeman}}, \bibinfo {author}
  {\bibfnamefont {K.~M.}\ \bibnamefont {Yu}}, \bibinfo {author} {\bibfnamefont
  {M.~S.}\ \bibnamefont {Brandt}}, \ and\ \bibinfo {author} {\bibfnamefont
  {O.~D.}\ \bibnamefont {Dubon}},\ }\href {\doibase 10.1103/PhysRevB.78.214421}
  {\bibfield  {journal} {\bibinfo  {journal} {Phys. Rev. B}\ }\textbf {\bibinfo
  {volume} {78}},\ \bibinfo {pages} {214421} (\bibinfo {year}
  {2008})}\BibitemShut {NoStop}%
\bibitem [{\citenamefont {Birss}\ and\ \citenamefont
  {Lee}(1960)}]{Birss:76:1960}%
  \BibitemOpen
  \bibfield  {author} {\bibinfo {author} {\bibfnamefont {R.~R.}\ \bibnamefont
  {Birss}}\ and\ \bibinfo {author} {\bibfnamefont {E.~W.}\ \bibnamefont
  {Lee}},\ }\href {\doibase 10.1088/0370-1328/76/4/307} {\bibfield  {journal}
  {\bibinfo  {journal} {Proc. Phys. Soc.}\ }\textbf {\bibinfo {volume} {76}},\
  \bibinfo {pages} {502} (\bibinfo {year} {1960})}\BibitemShut {NoStop}%
\bibitem [{\citenamefont {Alers}\ \emph {et~al.}(1960)\citenamefont {Alers},
  \citenamefont {Neighbours},\ and\ \citenamefont {Sato}}]{Alers:13:1960}%
  \BibitemOpen
  \bibfield  {author} {\bibinfo {author} {\bibfnamefont {G.~A.}\ \bibnamefont
  {Alers}}, \bibinfo {author} {\bibfnamefont {J.~R.}\ \bibnamefont
  {Neighbours}}, \ and\ \bibinfo {author} {\bibfnamefont {H.}~\bibnamefont
  {Sato}},\ }\href {\doibase 10.1016/0022-3697(60)90125-6} {\bibfield
  {journal} {\bibinfo  {journal} {J. Phys. Chem. Solids}\ }\textbf {\bibinfo
  {volume} {13}},\ \bibinfo {pages} {40 } (\bibinfo {year} {1960})}\BibitemShut
  {NoStop}%
\end{thebibliography}
\end{document}